\documentclass[12pt]{article}

\usepackage{epsf}
\newcommand{\bmat}{\left(\begin{array}}
\newcommand{\emat}{\end{array}\right)}
\def\NPB{Nucl. Phys. B}
\def\PLB{Phys. Lett. B}
\def\PLB{Phys. Lett. B}
\def\PRD{Phys. Rev. D}

\def\yzero{\smash{\hbox{$y\kern-4pt\raise1pt\hbox{${}^\circ$}$}}}

\def\a{\alpha}
\def\b{\beta}

\def\beq{\begin{equation}}
\def\eeq{\end{equation}}
\def\beqa{\begin{eqnarray}}
\def\eeqa{\end{eqnarray}}
\def\t{\times}

\def\th{\theta}

\def\-{\hphantom{-}}
\def\ov{\overline}
\def\s2{\frac{1}{\sqrt2}}

\def\oh{\frac{1}{2}}
\def\beq{\begin{equation}}
\def\eeq{\end{equation}}
\def\beqa{\begin{eqnarray}}
\def\eeqa{\end{eqnarray}}

\def\IF{\relax{\rm I\kern-.18em F}}
\def\II{\relax{\rm I\kern-.18em I}}
\def\IP{\relax{\rm I\kern-.18em P}}
\def\IC{\relax\hbox{\kern.25em$\inbar\kern-.3em{\rm C}$}}
\def\IR{\relax{\rm I\kern-.18em R}}

\def\cp{{\cal P}}

\def\Dsl{\,\raise.15ex\hbox{/}\mkern-13.5mu D} %this one can be subscripted
\def\IZ{Z\kern-.4em  Z}

 \def\cp#1{\relax\ifmmode {\IP\kern-2pt{}_{#1}}\else $\IP\kern-2pt{}_{#1}$\=fi}
\catcode`\@=11
\newdimen\@rotdimen
\newbox\@rotbox

\def\@vspec#1{\special{ps:#1}}%  passes #1 verbatim to the output
\def\@rotstart#1{\@vspec{gsave currentpoint currentpoint translate
   #1 neg exch neg exch translate}}% #1 can be any origin-fixing transformation
\def\@rotfinish{\@vspec{currentpoint grestore moveto}}% gets back in synch
\def\@rotr#1{\@rotdimen=\ht#1\advance\@rotdimen by\dp#1%
   \hbox to\@rotdimen{\hskip\ht#1\vbox to\wd#1{\@rotstart{90 rotate}%
   \box#1\vss}\hss}\@rotfinish}
\def\@rotl#1{\@rotdimen=\ht#1\advance\@rotdimen by\dp#1%
   \hbox to\@rotdimen{\vbox to\wd#1{\vskip\wd#1\@rotstart{270 rotate}%
   \box#1\vss}\hss}\@rotfinish}%
\def\@rotu#1{\@rotdimen=\ht#1\advance\@rotdimen by\dp#1%
   \hbox to\wd#1{\hskip\wd#1\vbox to\@rotdimen{\vskip\@rotdimen
   \@rotstart{-1 dup scale}\box#1\vss}\hss}\@rotfinish}%
\def\@rotf#1{\hbox to\wd#1{\hskip\wd#1\@rotstart{-1 1 scale}%
   \box#1\hss}\@rotfinish}%
\def\rotate{\@ifnextchar[{\@rotate}{\@rotate[l]}}
\def\@rotate[#1]#2{\setbox\@rotbox=\hbox{#2}\@nameuse{@rot#1}\@rotbox}

\catcode`\@=12
\begin{document}

%----------------------------------------------------------------------%
%  numbering equations with section number
%----------------------------------------------------------------------%
\makeatletter \@addtoreset{equation}{section} \makeatother
\renewcommand{\theequation}{\thesection.\arabic{equation}}
%----------------------------------------------------------------------%
%  title page
%----------------------------------------------------------------------%
\pagestyle{empty}
%\vspace{1.0cm}
%\rightline{FTUAM-1/27; IFT-UAM/CSIC-01-27}
%\vspace{2.5cm}
%----------------------------------------------------------------------%
%  Resetting of counters
%----------------------------------------------------------------------%
%\setcounter{page}{0}
\pagestyle{empty}
\vspace{0.5in}
\rightline{FTUAM-02/05}
\rightline{IFT-UAM/CSIC-02-03}
\rightline{\today}
\vspace{2.0cm}
\setcounter{footnote}{0}

\begin{center}
\LARGE{
{\bf GUT Model Hierarchies from Intersecting Branes}}
\\[4mm]
%\medskip
{\large{ Christos ~Kokorelis \footnote{ Christos.Kokorelis@uam.es} }
\\[1mm]}
\normalsize{\em Departamento de F\'\i sica Te\'orica C-XI and 
Instituto de F\'\i sica 
Te\'orica C-XVI}
,\\[-0.3em]
{\em Universidad Aut\'onoma de Madrid, Cantoblanco, 28049, Madrid, Spain}
\end{center}
\vspace{1.0mm}

%%%%%%%%%%%%%%%%%%%%%%%%%%%%%%%%%%%%%%%

%\begin{center}
%\begin{minipage}[h]{14.5cm}
\begin{center}
{\small  ABSTRACT}
\end{center}
By employing D6-branes
intersecting at angles in $D = 4$ type I strings, 
we construct the first examples of three generation string GUT
models (PS-A class), that contain at low energy exactly the
standard model spectrum with no extra matter and/or
extra gauge group factors.
They are based on the group
$SU(4)_C \times SU(2)_L \times SU(2)_R$.
The models are non-supersymmetric, even though
SUSY is 
unbroken in the bulk.
Baryon number is gauged  
and its anomalies are cancelled through a generalized Green-Schwarz
mechanism. We also discuss models (PS-B class) which at low energy have the
standard model
augmented by an anomaly free $U(1)$ symmetry and 
show that multibrane wrappings correspond to a trivial  
redefinition of the surviving global $U(1)$ at low energies.
There are no colour triplet couplings to mediate proton decay
 and proton is stable.
The models are compatible with a low string scale of energy less that
650 GeV and are
directly testable at present or future accelerators as
they predict the existence of light left handed weak fermion doublets at
energies between 90 and 246 GeV.
The neutrinos get a mass through an unconventional see-saw mechanism.
The mass relation $m_e = m_d$ at the GUT scale is
recovered.
Imposing supersymmetry at particular intersections generates
non-zero Majorana masses for right handed neutrinos as well providing the
necessary singlets needed to break the surviving anomaly free $U(1)$, thus
suggesting a gauge symmetry breaking method that can be applied in 
general left-right symmetric models.
%\end{minipage}                 
%\end{center}

\newpage
%----------------------------------------------------------------------%
%  Resetting of counters
%----------------------------------------------------------------------%
\setcounter{page}{1} \pagestyle{plain}
\renewcommand{\thefootnote}{\arabic{footnote}}
\setcounter{footnote}{0}
%----------------------------------------------------------------------%
%  Paper begins
%----------------------------------------------------------------------%

\section{Introduction}

While string theory remains the only candidate for a consistent theory
of fundamental interactions it still has to solve some major problems like 
explaining the hierarchy of scale and particle masses after 
supersymmetry breaking. 
These phenomelogical issues have by far been explored in the context of 
construction of semirealistic supersymmetric models 
of 
weakly coupled heterotic string theories \cite{ena}.
Leaving aside the weakly coupled heterotic string,
$N=1$ four-dimensional orientifold models \cite{dio1} represent a particular 
class of consistent string solutions which explore the physics of strongly
coupled heterotic strings.
Semirealistic model building has been explored in the context of
$N=1$ supersymmetric (SUSY) four-dimensional orientifolds \cite{ibaba}. 
The main futures of the models
constructed include an extended gauge group which includes the standard model
or extensions of it, with a variety of exotic matter.

Recently some new constructions have appeared in a type I 
string vacuum background which use 
intersecting branes \cite{tessera} and give four dimensional
non-supersymmetric models. These are the kind of models that we
will be examining in this work.  The question that someone
might address at this point is why we have to use non-SUSY models while
in heterotic string compactifications we examined SUSY one's?
The reason for doing so is mainly phenomenologigal.
In $N=1$ (orbifold) compactifications of the heterotic string
the string scale was of the order of $10^{18}$ GeV something that
was in clear disagreement with the observed unification of gauge
coupling constants in the MSSM of $10^{16}$ GeV.
In these models the observed discrepancy between the two high 
scales was attributed to the presence of the $N=1$
string threshold corrections
to the gauge coupling constants \cite{dikomou}.
On the contrary in type I models the string scale is a free parameter.
Moreover,  
recent results suggest
the string scale in type I models can be in the TeV range
\cite{antoba}. The latter result suggests that
non-SUSY models with a string scale in the TeV region
is a viable possibility.

Because in the open string models of \cite{tessera}
background fluxes were used,
following past ideas about the use of 
magnetic fields in open strings \cite{pente}, 
in a D9 brane type I background with background
fluxes \footnote{
In the T-dual 
language these backgrounds are represented by
D6 branes wrapping 3-cycles on a dual torus
and intersecting each other at certain angles. }.
it was possible to break supersymmetry on the 
brane and to get chiral fermions 
with an even number of 
generations \cite{tessera}.
The fermions on those models appear in the intersections between
branes \cite{tessera1}, \cite{bele}.

After introducing a quantized background NS-NS B 
field \cite{eksi1,eksi2,eksi3}, that makes the tori tilted, is was then
possible to get 
semirealistic models with 
three generations \cite{tessera2}. We also note that these backgrounds
are T-dual to models with magnetic deformations \cite{carlo}.

Additional non-SUSY constructions in the context 
of intersecting branes,
from IIB orientifolds, consisting of getting at 
low energy the standard model 
spectrum with extra matter and additional chiral fermions
were derived in \cite{luis1}.
The construction involves
D(3+n) branes wrapping on the compact space $T^{2n} \times 
(T^{2(3-n)}/Z_N)$, for $n=1, 2, 3$ and intersecting at angles
in the $T^{2n}$.

Furthermore, an important step was taken in \cite{louis2},
by showing how to construct
the standard model (SM) spectrum together with right handed 
neutrinos in a systematic way. The authors considered, as a
starting point,
IIA theory compactified on $T^6$ assigned with an orientifold  
product $\Omega \times R$, where $\Omega$ is the worldsheet parity
operator
and R is the reflection operator with respect to one of the axis of
each tori.
In this case, the four stack D6-branes contain Minkowski space and  
each of the 
three remaining dimensions is wrapped up on a different $T^2$ torus.
In this
construction the proton is stable since the baryon number is a gauged
$U(1)$ global symmetry.
A special feature of these models is that the neutrinos can only 
get Dirac mass. These models have been generalized to models with
five stack
of D6-branes at \cite{kokos}.
 For a discussion of non-SUSY SM in the context of 
D3-branes on orbifold singularities see \cite{alda}.  
  A different attempt to construct non-SUSY GUT models in the context of 
intersecting
branes was made in \cite{blume}.
However, there were some problems with the phenomenology of the 
$SU(5)$ GUT model presented, as some of the 
Yukawa couplings
were excluded
and the standard electroweak Higgs scalar was not realized,
while proton decay problems appeared. Also SUSY constructions
in the context of intersecting branes were considered in
\cite{uran}.
Nevertheless, despite the fact that much progress has been made,
constructing string models with interesting
phenomenology is still a difficult task.

 The purpose of this paper is 
to
present the first three generation string models that are based on
a grand unified gauge group, and contain at low energy exactly
the standard model spectrum, namely $SU(3)_C \t SU(2)_L \t U(1)_Y$,
without any extra chiral fermions and/or extra gauge group factors.  
The four-dimensional models are non-supersymmetric intersecting brane 
constructions and are based on the Pati-Salam (PS)
$SU(4)_C \times SU(2)_L \times SU(2)_R$ gauge group \cite{Pati-Salam}.
The basic structure behind the models includes
D6-branes intersecting each other at non-trivial angles,
in an orientifolded
factorized six-torus,
where O$_6$ orientifold planes are on top of D6-branes.  
\newline
The proposed models have some distinctive features :
\begin{itemize}

\item   The models (characterized as belonging to the PS-A class)
start with a
gauge group at the string
scale $U(4) \times
U(2) \times U(2) \times U(1)$. At the scale of symmetry
breaking of the left-right symmetry, $M_{GUT}$, the initial symmetry group
breaks to the the standard model  
$SU(3)_C \times SU(2)_L \times U(1)_Y $
 augmented with an extra anomaly free $U(1)$ symmetry.
 The additional $U(1)$ symmetry breaks by the vev of charged singlet
 scalars, e.g. $ s_L^H$, to the SM itself at a scale set by its vev.
  The singlets responsible for breaking the  $U(1)$ symmetry are obtained by
 demanding certain open string sectors of the non-SUSY model to respect
 $N=1$ supersymmetry.

\item  Neutrinos gets a mass 
   of the right order, consistent with the LSND oscillation 
experiments \cite{agui}, 
from a see-saw mechanism of the Frogatt-Nielsen 
type \cite{fro}.
The structure of Yukawa couplings involved in the see-saw 
mechanism \cite{raya} supports the smalleness of neutrino masses thus
generating a hierarchy in consistency with neutrino oscillation
experiments.

\item  Proton is stable due to the fact that baryon number is an 
unbroken gauged 
global symmetry surviving at low energies and no colour triplet
couplings that could mediate
proton decay exist. 
Thus a gauged baryon number provides a natural explanation for 
proton stability.
As in the models of \cite{louis2}
the baryon number associated
$U(1)$ gauge boson becomes massive through its couplings to Green-Schwarz
mechanism. That has an an immediate effect that baryon number is surviving
as a global symmetry to low energies providing for a natural explanation
for proton stability in general brane-world scenarios.

\item 
The model uses small Higgs representations in the adjoint
to break the PS symmetry,
instead of using large
Higgs representations, e.g. 126 like in the standard $SO(10)$ models.

\item
The bidoublet Higgs fields $h$ responsible for electroweak symmetry breaking
do not get charged under the global $U(1)$ and thus lepton number is 
not broken at the standard model.

\end{itemize}

We should note that in the past three generation four dimensional 
string vacua that
include the 
$SU(4)_C \times SU(2)_L \times SU(2)_R$ gauge group together with
extra matter
and additional non-abelian gauge group structure have been discussed
both in the context of supersymmetric vacua coming from orientifolds of type 
IIB \cite{kaku}
and from non-supersymmetric brane-antibrane 
pair configurations \cite{luis3}. 
For some other proposals for realistic D-brane model building, 
based not 
on a particular string construction, see \cite{antokt}
for the standard model, \cite{lr} for the PS model or 
for the
standard model in a non-compact set-up \cite{bere}.

The paper is organized as follows. 
 In chapter two we describe
the general characteristics of the models, with
particular emphasis on
how to 
calculate the fermionic spectrum from intersecting branes, as well providing 
the
multi-parameter solutions to the RR tadpole cancellation conditions.
We discuss two kinds of models, characterized in this work as
belonging to the PS-A, PS-B classes of models.
In addition we discuss how the PS-A classes of models 
accommodate singlet fields. The latter fields are
necessary in order to break the
surviving $U(1)$ symmetry and
getting just the SM at low energy.
In chapter 3 we examine the cancellation of $U(1)$ anomalies via a 
generalized 
Green-Schwarz (GS) mechanism finding the general solution 
for the non-anomalous $U(1)$
which remains light. We also discuss arguments related to
 multiwrapping branes
and show that they correspond to a trivial redefinition of the global
non-anomalous $U(1)$ surviving the GS mechanism.  
In chapter 4 we discuss the Higgs sector of the model involving
the appearance of Higgs scalar responsible for breaking the
PS $SU(4) \otimes SU(2)_R$
symmetry at the intermediate grand unified scale $M_{GUT}$ and the
electroweak breaking Higgs scalars.
We also discuss how the imposition of supersymmetry in particular
sectors of the classes of models succeeds to break the PS-A class to the
SM itself at low energies, even though there is not a similar effect for
the PS-B class of models.
In chapter 5 we examine the problem of
neutrino masses. We also show that for the PS-A class of models all
additional fermions beyond those of SM become massive
and disappear from the low energy spectrum.
In this section, we describe in detail how the presence of supersymmetry
in particular sectors of the theory realizes the particular couplings
taking part in the see-saw mechanism. We also discuss
bounds for the string scale and right handed neutrino masses that
follow from the Yukawa couplings of the models.
Chapter 6 contains our conclusions. Finally, 
Appendices I, II include the conditions for the absence of tachyonic modes
in the spectrum of the PS-A, PS-B class of models presented,
while in Appendix III
we provide an equivalent structure to PS-B class of models
presented in the main boby of this article together with its 
tadpole solutions.

\section{The models and the rules of computing the spectrum}

In the present work, we are going to look for a three
family non-supersymmetric
model that is based on the left-right
symmetric  $SU(4)_C \times SU(2)_L \times SU(2)_R$
Pati-Salam model with the right
phenomenological
properties and discuss in more detail its phenomenology.
It will
come from D6-branes wrapping on 3-cycles of 
toroidal orientifolds of type I in four dimensions.
We will preseent a simultaneous discussion of the two classes of
PS models, PS-A and PS-B, so unless otherwise stated the discussions
will hold for both classes of models.
Let at this point describe 
the general futures of the non-supersymmetric  
$SU(4)_C \times SU(2)_L \times SU(2)_R$  
model. 
Important characteristic of all vacua coming from these type I 
constructions is the replication of massless fermion spectrum
by an equal number of massive particles in the same representations 
and with the same quantum numbers.\newline
The quark and lepton 
fields appear in three complete generations and 
are accommodated into the following representations : 
\beqa
F_L &=& (4, {\bar 2}, 1) =\ q(3, {\bar 2}, \frac{1}{6})
+\ l(1, {\bar 2}, -\frac{1}{2})
\equiv\ (\ u,\ d,\ l), \nonumber\\
{\bar F}_R &=& ({\bar 4}, 1,  2) =\ {u}^c({\bar 3}, 1, - \frac{2}{3}) +
{d}^c({\bar 3}, 1, \frac{1}{3}) + {e}^c(1, 1, 1) + {N}^c(1, 1, 0) 
\equiv ( {u}^c,  {d}^c,   {l}^c),\nonumber\\ 
\label{na3}
\eeqa
where the quantum numbers on the right hand side of 
(\ref{na3})
are with respect to the decomposition of the $SU(4)_C \times SU(2)_L \times 
SU(2)_R$ under the $SU(3)_C \times SU(2)_L \times U(1)_Y$ gauge group and
${l}=(\nu,e) $ is the standard left handed lepton doublet,
 ${l}^c=({N}^c, e^c)$ are the right handed leptons.
Note that the assignment of the accommodation of the quarks and leptons into
the representations $F_L + {\bar F}_R$ is the one appearing in the
spinorial decomposition of the $16$ representation of $SO(10)$ under the PS 
gauge group.
\newline
A set of useful fermions appear also in the model
\beq
\chi_L =\ ( 1,  {\bar 2}, 1),\ \chi_R =\ (1, 1, {\bar 2}).
\label{useful}
\eeq
These fermions are a general prediction
of left-right symmetric theories.
As we comment later the existence of these representations in the model
follows from RR tadpole cancellation conditions. 
\newline
The symmetry breaking of the left-right PS symmetry at the $M_{GUT}$
breaking scale \footnote{In principle this scale can be lower than the
string scale.}
proceeds through the representations of the set of Higgs fields,
\beq
H_1 =\ ({\bar 4}, 1, {\bar 2}), \
H_2 =\ ( 4, 1, 2),
\label{useful1}
\eeq
 where, e.g.
\beqa
H_1 = ({\bar 4}, 1, {\bar 2}) = {u}_H({\bar 3}, 1, \frac{2}{3}) +
{d}_H({\bar 3}, 1, -\frac{1}{3}) + {e}_H(1, 1, -1)+
{\nu}_H(1, 1, 0).
\label{higgs1}
\eeqa
The electroweak symmetry is delivered 
through bi-doublet Higgs fields $h_i$ $i=1,2$,
field in the 
representations 
\beq
h_1 =\ (1, 2, 2),\ h_2 =\ (1, {\bar 2}, {\bar 2}) \ .
\label{additi1}
\eeq
Also present are the massive scalar superpartners
\footnote{These fields are replicas
of the fermion fields appearing in the intersection 
$ac$ of table one.}
of the quarks, leptons and
antiparticles 
\beq
{\bar F}_R^H = ({\bar 4}, 1,  2) = {u}^c_H({\bar 3}, 1, -\frac{4}{6})+
{d}^c_H({\bar 3}, 1, \frac{1}{3})+ {e}^c_H(1, 1, 1) + 
{N}^c_H(1, 1, 0) 
\equiv ({u}^c_H, {d}^c_H,  {l}^c_H).\nonumber\\
\label{na368}
\eeq
Also, only for the PS-A class models, a number of charged exotic
fermion fields appear
\beq
12(6, 1, 1),\;\;6({\bar 6}, 1, 1),\;\;
6({\bar 10}, 1, 1),\;\;
24(1, 1, 1, 1)
\label{beg1}
\eeq
as well as the singlets
\beq
24(1, 1, 1)^H
\label{beg2}
\eeq

Next, we describe the construction of the PS classes of models. It is 
based on  
type I string with D9-branes compactified on a six-dimensional
orientifolded torus $T^6$, 
where internal background 
gauge fluxes on the branes are turned on. 
By performing a T-duality transformation on the $x^4$, $x^5$, $x^6$, 
directions the D9-branes with fluxes are translated into D6-branes 
intersecting at 
angles. The branes are not paralled to the
orientifold planes.
We assume that the D$6_a$-branes are wrapping 1-cycles 
$(n^i_a, m^i_a)$ along each of the $T^2$
torus of the factorized $T^6$ torus, namely 
$T^6 = T^2 \times T^2 \times T^2$.

In order to build a PS model with minimal Higgs structure we consider
four stacks of D6-branes giving rise to their world-volume to an initial
gauge group $U(4) \times U(2) \times U(2) \times U(1)$ at the string
scale.
In addition, we consider the addition of NS B-flux, 
such that the tori are not orthogonal,
avoiding in this way an even number of families, 
and leading to effective tilted wrapping numbers, 
\beq
(n^i, m ={\tilde m}^i + {n^i}/2);\; n,\;{\tilde m}\;\in\;Z,
\label{na2}
\eeq
that allows semi-integer values for the m-numbers.  
\newline
Because of the $\Omega {\cal R}$ symmetry, 
where $\Omega$ is the worldvolume 
parity and $\cal R$ is the reflection on the T-dualized coordinates,
\beq
T (\Omega {\cal R})T^{-1}= \Omega {\cal R},
\label{dual}
\eeq
 each D$6_a$-brane
1-cycle, must have its $\Omega {\cal R}$ partner $(n^i_a, -m^i_a)$. 

Chiral fermions are obtained by stretched open strings between
intersecting D6-branes \cite{bele}. 
The chiral spectrum of the models is obtained after solving simultaneously 
the intersection 
constraints coming from the existence of the different sectors together
with the RR
tadpole cancellation conditions.

There are a number of different sectors, 
which should be taken into account when computing the chiral spectrum.
We denote the action of 
$\Omega R$ on a sector $\a, \b$, by ${\a}^{\star}, {\b}^{\star}$,
respectively.
The possible sectors are:

\begin{itemize}
 
\item The $\a \b + \b \a$ sector: involves open strings stretching
between the
D$6_{\a}$ and D$6_{\b}$ branes. Under the $\Omega R$ symmetry 
this sector is mapped to its image, ${\a}^{\star} {\b}^{\star}
+ {\b}^{\star} {\a}^{\star}$ sector.
The number, $I_{\a\b}$, of chiral fermions in this sector, 
transforms in the bifundamental representation
$(N_{\a}, {\bar N}_{\a})$ of $U(N_{\a}) \times U(N_{\b})$, and reads
\beq
I_{\a\b} = ( n_{\a}^1 m_{\b}^1 - m_{\a}^1 n_{\b}^1)( n_{\a}^2 m_{\b}^2 -
m_{\a}^2 n_{\b}^2 )
(n_{\a}^3 m_{\b}^3 - m_{\a}^3 n_{\b}^3),
\label{ena3}
\eeq
 where $I_{\a\b}$
is the intersection number of the wrapped cycles. Note that the sign of
 $I_{\a\b}$ denotes the chirality of the fermion and with $I_{\a\b} > 0$
we denote left handed fermions.
Negative multiplicity denotes opposite chirality.

\item The $\a\a$ sector : it involves open strings stretching on a single 
stack of 
D$6_{\a}$ branes.  Under the $\Omega R$ symmetry 
this sector is mapped to its image ${\a}^{\star}{\a}^{\star}$.
 This sector contain ${\cal N}=4$ super Yang Yills and if it exists
SO(N), SP(N) groups appear. 
This sector is of no importance to us as we are 
interested in unitary groups.

\item The ${\a} {\b}^{\star} + {\b}^{\star} {\a}$ sector :
It involves chiral fermions transforming into the $(N_{\a}, N_{\b})$
representation with multiplicity given by
\beq
I_{{\a}{\b}^{\star}} = -( n_{\a}^1 m_{\b}^1 + m_{\a}^1 n_{\b}^1)
( n_{\a}^2 m_{\b}^2 + m_{\a}^2 n_{\b}^2 )
(n_{\a}^3 m_{\b}^3 + m_{\a}^3 n_{\b}^3).
\label{ena31}
\eeq
Under the $\Omega R$ symmetry transforms to itself.

\item the ${\a} {\a}^{\star}$ sector : under the $\Omega R$ symmetry is 
transformed to itself. From this sector the invariant intersections
will give 8$m_{\a}^1 m_{\a}^2 m_{\a}^3$ fermions in the
antisymmetric representation
and the non-invariant intersections that come in pairs provide us with
4$ m_{\a}^1 m_{\a}^2 m_{\a}^3 (n_{\a}^1 n_{\a}^2 n_{\a}^3 -1)$ additional 
fermions in the symmetric and 
antisymmetric representation of the $U(N_{\a})$ gauge group.

\end{itemize}

Any vacuum derived from the previous intersection number constraints of the 
chiral spectrum 
is subject to constraints coming from RR tadpole cancellation 
conditions \cite{tessera}. That requires cancellation of 
D6-branes charges \footnote{Taken together with their
orientifold images $(n_a^i, - m_a^i)$  wrapping
on three cycles of homology
class $[\Pi_{\alpha^{\star}}]$.}, wrapping on three cycles with
homology $[\Pi_a]$ and O6-plane 7-form
charges wrapping on 3-cycles with homology $[\Pi_{O_6}]$. In formal terms,
the RR tadpole cancellation  conditions
in terms of cancellations of RR charges in homology, read :
\beq
\sum_a N_a [\Pi_a]+\sum_{\alpha^{\star}} 
N_{\alpha^{\star}}[\Pi_{\alpha^{\star}}] -32
[\Pi_{O_6}]=0.
\label{homology}
\eeq  
Explicitly, the RR tadpole conditions read :
\beqa
\sum_a N_a n_a^1 n_a^2 n_a^3 =\ 16,\nonumber\\
\sum_a N_a m_a^1 m_a^2 n_a^3 =\ 0,\nonumber\\
\sum_a N_a m_a^1 n_a^2 m_a^3 =\ 0,\nonumber\\
\sum_a N_a n_a^1 m_a^2 m_a^3 =\ 0.
\label{na1}
\eeqa
That ensures absense of non-abelian gauge anomalies.
A comment is in order. It is important to notice
that the RR tadpole cancellation condition can be understood as
a constraint that demands that for each gauge group the number of
fundamentals to be equal to the number of bifundamenals.
As a general rule to D-brane model building, by considering $a$ 
stacks of D-brane configurations with 
$N_a, a=1, \cdots, N$, paralled branes, the gauge group appearing is in 
the form $U(N_1) \times U(N_2) \times \cdots \times U(N_a)$. 
Effectively, each $U(N_i)$ factor will give rise to an $SU(N_i)$
charged under the associated $U(1_i)$ gauge group factor that appears in 
the decomposition $SU(N_a) \times U(1_a)$.
A type I brane configuration with the unique minimal PS particle content
such that intersection numbers, tadpole conditions and various phenomenological
requirements including the absence of exotic representations are accommodated, 
 can be 
obtained by considering four stacks of branes yielding an initial
$U(4)_a \times U(2)_b \times U(2)_c \times U(1)_d $ gauge group equivalent
to an $SU(4)_a \times SU(2)_b \times SU(2)_b \times U(1)_a \times U(1)_b 
\times U(1)_c \times U(1)_d $. Thus, in the first instance, 
we can identity, without loss of 
generality, $SU(4)_a$ as the $SU(4)_c$ colour group 
that its breaking could induce 
the usual $SU(3)$ colour group of strong interactions, the $SU(2)_b$ with 
$SU(2)_L$ of weak interactions and $SU(2)_c$ with $SU(2)_R$.

The complete accommodation of the fermion structure of the model under
study can be seen
in table one.

\begin{table}[htb] \footnotesize
\renewcommand{\arraystretch}{1.5}
\begin{center}
\begin{tabular}{|c|c||c|c||c||c|c|}
\hline
Fields &Intersection  & $\bullet$ $SU(4)_C \times SU(2)_L \times SU(2)_R$
 $\bullet$&
$Q_a$ & $Q_b$ & $Q_c$ & $Q_d$ \\
\hline
 $F_L$& $I_{ab^{\ast}}=3$ &
$3 \times (4,  2, 1)$ & $1$ & $1$ & $0$ &$0$ \\
 ${\bar F}_R$  &$I_{a c}=-3 $ & $3 \times ({\ov 4}, 1, 2)$ &
$-1$ & $0$ & $1$ & $0$\\
 $\chi_L$& $I_{bd} = -12$ &  $12 \times (1, {\ov 2}, 1)$ &
$0$ & $-1$ & $0$ & $1$ \\    
 $\chi_R$& $I_{cd^{\ast}} = -12$ &  $12 \times (1, 1, {\ov 2})$ &
$0$ & $0$ & $-1$ &$-1$ \\\hline
 $\omega_L$& $I_{aa^{\ast}}$ &  $12 \b^2 {\tilde \epsilon} \times (6, 1, 1)$ &
$2{\tilde \epsilon}$ & $0$ & $0$ &$0$ \\
 $z_R$& $I_{aa^{\ast}}$ & $6  \b^2 {\tilde \epsilon}  \times ({\bar 10}, 1, 1)$ &
$-2{\tilde \epsilon}$ & $0$ & $0$ &$0$ \\
 $s_L$& $I_{dd^{\ast}}$ &  $24 \b^2 {\tilde \epsilon} \times (1, 1, 1)$ &
$0$ & $0$ & $0$ &$-2{\tilde \epsilon}$ \\
\hline
\end{tabular}
\end{center}
\caption{\small Fermionic spectrum of the $SU(4)_C \times
SU(2)_L \times SU(2)_R$, type I models together with $U(1)$ charges.
The spectrum appearing in the full table is of
PS-A models. The top part corresponds to PS-B models. Note that the
representation context in the bottom part is considered by assuming
${\tilde \epsilon} = 1$. In the general case
${\tilde \epsilon} = \pm 1$. If ${\tilde \epsilon} = -1$ then the conjugate
fields should be considered, e.g. if ${\tilde \epsilon} = -1$, the
$\omega_L$ field should transform as $({\bar 6}, 1, 1)_{(-2, 0, 0, 0)}$.
\label{spectrum8}}
\end{table}

We note a number of interesting comments :
\newline
a)Two main directions towards model building classes of PS-models will be 
emphasized in this work. We can either choose to include
sectors $\a \a^{\ast}$ in the model, we call this class of models type PS-A
or not to include them, we call this class of models type PS-B. 
In the former case, PS-A,  the surviving gauge
group at low energies is exactly the SM. We get, at low energies, just
the fermionic content of the SM spectrum with all particles having
the correct hypercharge assignment.
The fermionic spectrum of PS-A models is given by the full spectrum 
appearing in table
(\ref{spectrum8}). The tadpole solutions in this case appear in table
(\ref{spectruma101}).
In the latter case, PS-B classes of models, the gauge group at
low energies is the SM augmented by an extra anomaly free $U(1)$.
The tadpole solutions for PS-B models appear in table 
(\ref{spectrum10}).   
The fermionic particle content of PS-B models appear in the four top rows of table
(\ref{spectrum8}).

Also, in order to
realize certain couplings we will impose that some intersections will
preserve some supersymmetry. In both PS-A, PS-B models, some massive
fields will be
``pulled out" from the massive spectrum and become massless.
For example, in order to realize a Majorana mass term for the right
handed neutrinos for both PS-A, PS-B models we will demand that the sector
$ac$ preserves $N=1$ SUSY. That will have as an immediate effect
to "pull out" from the massive mode spectrum the $F_R^H$ particles. 
\newline
b)
The intersection numbers, in table one,  
of the fermions $F_L + {\bar F}_R$ are chosen 
such that $I_{ac} = - 3$, $I_{ab} = 3$. Here, $-3$ denotes opposite 
chirality to that of a left handed fermion. 
The choise of additional fermion fields $(1, {\bar 2} ,1)$, 
$(1, 1, {\bar 2})$ is imposed to us by the RR tadpole cancellation conditions
that are equivalent to
$SU(N_a)$ gauge
anomaly cancellation, in this case of $SU(2)_L$, $SU(2)_R$ gauge anomalies,
\beq
\sum_i I_{ia} N_a = 0,\;\;a = L, R.
\label{ena4}
\eeq
c) The PS-A, PS-B classes of models lack representations
of scalar sextets $(6, 1, 1)$ fields,
that appear in attempts to construct realistic 4D $N=1$ PS heterotic models 
from
the fermionic formulation \cite{anto} or in  
D-brane inspired models \cite{lr},
even 
through examples of heterotic fermionic models where those 
representations are lacking 
exist \cite{giapo}.
Those representations were imposed earlier in attempts to produce a
realistic PS model \footnote{See the first reference of \cite{anto}.}
as a recipe for saving the models from proton decay.
Fast proton decay was avoided by making 
the mediating 
$d_H$ triplets of (\ref{higgs1}) superheavy 
and of order of the $SU(2)_R$ breaking scale
via their couplings to the sextets.
In the models we examine in this work,
baryon number is a gauged global symmetry,
so that proton is 
stable. Thus there is no need to introduce sextets to save the models
from fact proton decay as proton is stable anyhow.
\newline
Also in this case, there is no problem of having $d_H$ becoming light
enough and causing catastrofic proton decay, as the only way this could
happen, is through the existense of the $d_H$ coupling to sextets 
to quarks and leptons. But such a coupling is forbidden by
the symmetries 
of the models by construction.
\newline
Also, the PS-B model class has some shortcomings. 
The weak and right doublets
$\chi_L$, $\chi_R$ respectively survive massless
at low energies of order $M_Z$. Both massless
particles are unwelcome as they are not observed at energies of
order $M_Z$.
Nevertheless this case is interesting as a number of 
useful conclusions could be derived from the study of those models.
\newline
To be convinced that scalar sextet fields cannot exist in intersecting
PS-B type I
D-brane models let us imagine that they do existed \footnote{Introducing
scalar sextets in this case would demand imposing $N=1$ SUSY in this sector
such that the full $N=1$ sextet hypermultiplet would be massless.}.
Then it may then be easily seen that with four 
stacks of branes, they would have to be \footnote{
An alternative equivalent choise of 
$(6, 1, 1)_{(\;1,0, 0,\;-1)}$, $({\bar 6}, 1, 1)_{(-1, 0, 0, 1)}$
would demand $I_{ad} = 1$, $I_{ad} = -1$ which is impossible anyway to 
accommodate.
Even by using a PS-B model
with five stacks of branes, or more, 
e.g. an $U(4) \t U(2) \t U(2) \t U(1) \t U(1)$, it will be
impossible to accommodate sextet fields like those in  (\ref{ena5})
for similar reasons.} in the form :

\beq
(6, 1, 1)_{(\;1,0, 0,\; 1)},\;({\bar 6}, 1, 1)_{(-1, 0, 0, -1)}.
\label{ena5}
\eeq
This choise is consistent with the cancellation of mixed 
anomalies of $U(1)$'s
with the non-abelian gauge group factors. However, 
this choise demands
\beq
I_{ad^{\star}} = -1\;for\;(6, 1, 1);\;I_{ad^{\star}} = 1\;for\;
({\bar 6}, 1, 1).
\label{ena6}
\eeq
Obviously, it is not possible to accommodate simultaneously 
the two different intersection numbers in (\ref{ena6}), ruling out 
the problematic representations (\ref{ena5}).

For PS-A classes of models, there are no
shortcomings. The theory breaks
just to the standard model $SU(3) \times SU(2) \times U(1)_Y$
at low energies.
 The complete spectrum
of the model appears in table (\ref{spectrum10}). The tadpole solutions
of PS-A
models are presented in table (\ref{spectruma101}).
\newline
d) The mixed anomalies $A_{ij}$ of the four surplus $U(1)$'s 
with the non-abelian gauge groups $SU(N_a)$ of the theory
cancel through a generalized GS mechanism \cite{sa,iru},
involving
 close string modes couplings to worldsheet gauge fields.
 Two combinations of the $U(1)$'s are anomalous and become
 massive through their
 couplings to RR fields, their
 orthogonal 
 non-anomalous combinations survives, combining to a single $U(1)$
 that remains massless.
\newline
e)
For PS-A models the constraint 
\beqa
{\Pi}_{i=1}^3 m^i&=&0.
\label{req1}
\eeqa                  
is not imposed 
and thus
leads to the appearance of the non-trivial chiral fermion content from the
$aa^{\ast}$, $dd^{\ast}$ sectors with corresponding fermions $\omega_L$,
$y_R$, $z_R$, $s_Z$.
After breaking the PS left-right symmetry at $M_{GUT}$, the
surviving gauge symmetry
is that of the SM augmented by an anomaly free $U(1)$ symmetry
surviving the Green-Schwarz mechanism.  
To break the latter $U(1)$ symmetry we will impose
that the  $dd^{\star}$ sector respects $N=1$ SUSY.
Thus singlets scalars will appear, that are superpartners of $s_L$
fermions.

 For type PS-B models, in order to cancel the appearance of
exotic representations in the model
appearing from the general $D D^{\star}$ sectors, 
in antisymmetric and symmetric 
representations of the $U(N_a)$ group, we require that 
(\ref{req1}) constraint 
holds \footnote{The same constraint was working perfectly at the 
level of building 
just the 
standard model at low energies, starting from stacks of branes that are
not based on
a non-GUT group at the string scale.
For examle see \cite{louis2} for the four-stack D6 SM and \cite{kokos} for the
five stack D6 SM.}.

Note that the choise of fermion fields for PS-B models
in table (\ref{spectrum10}) is
absolutely minimal, as 
a different choise of the set of fields with
three stacks of branes,
does not have a tadpole solution as long as we demand (\ref{req1}). 
\newline
f) Demanding $I_{ab}=3$, $I_{ac}=-3$,
it implies that the third tori should be tilted. By looking at the 
intersection numbers of table one,  we conclude that the
b-brane should be paralled to the c-brane and the a-brane should be 
paralled to the d-brane as there is an absence of intersection numbers for 
those branes. The complete list of intersection numbers for PS-B class is 
listed in table two.

\begin{table} [htb] \footnotesize
\renewcommand{\arraystretch}{2}
\begin{center}
\begin{tabular}{|c||c|c||c||c|c|}
\hline
  $I_{ab^{\star}}=3$ &
$I_{ac}=-3$ & $I_{bd}=-12$ & $I_{cd^{\ast}}=-12$ & $I_{ad}=0$ & $I_{bc}=0$ \\
\hline$I_{ab}=0$ &
$I_{ac^{\ast}}=0$ & $I_{bd^{\ast}}=0$ & $I_{cd}=0$ & $I_{ad^{\ast}}=0$ 
& $I_{bc^{\ast}}=0$ \\
\hline\hline
\end{tabular}
\end{center}
\caption{\small List of intersection constraints for the $SU(4)_C \times 
SU(2)_L \times SU(2)_R$ type I PS-B classes of models.
\label{spectrum9}}
\end{table}

The  
cancellation of the RR crosscap tadpole constraints
is solved from parametric sets of solutions.
For PS-A and PS-B classes of models they are given in tables
(\ref{spectruma101}) and (\ref{spectrum10}) respectively.

\begin{itemize}

\item {\em Tadpoles for PS-A classes of models}

For the PS-A classes of models, giving just the SM at low energies,
the choise of wrapping numbers appearing in table (\ref{spectruma101})
satisfies all tadpole conditions but the third of eqn's (\ref{na1}).
The latter becomes
\beq
2n_a^2+n_d^2+\frac{1}{ \b^2}(m_b^1-m_c^1)=0,
\label{psatad1}
\eeq
which may be solved by either
\beq
n_d^2 = \ -2n_a^2, \ \;m_b^1 = \ m_c^1
\label{no1} 
\eeq
or
\beq
2n_a^2 = \ \frac{m_c^1}{\b_2},\; \ n_d^2=-\frac{m_b^1}{\b_2},
\label{no2} 
\eeq
or
\beq
2n_a^2 = \ - \frac{m_b^1}{\b_2}, \  \;n_d^2=\frac{m_c^1}{\b_2}
\label{no3} 
\eeq
Choosing for example the solutions (\ref{no1}) we are
effectively making the tadpole solutions of table (\ref{spectruma101}) to
depend on one integer $n_a^2$ and the
tilted wrapping number $m_b^1$, the phase parameters
$\epsilon = \pm 1$,
$\tilde \epsilon = \pm 1$ and the NS-background
parameter
$\beta_i =\ 1-b_i$, which is associated to the parametrization of the 
NS B-field with $b_i =0,\ 1/2$.
In this case, an example of wrappings satisfying all tadpoles is given
by the choise of
 wrappings
\beq
n_a^2 =-1,\;m_b^1 = m_c^1 =1,\;\b_1=1,\;\b_2=1/2,\;{\tilde \epsilon}= -1.
\label{were}
\eeq
\begin{center}
\beqa
N_a =4&(0, \epsilon)(-1, 3/2)(-1,  -1/2) \nonumber\\
N_b =2&(-1, \  \ \epsilon )(2, \ 0) (-1, \ -1/2) \nonumber\\
N_c =2&(1,\  \ \epsilon )(2,\ 0)(-1, 1/2) \nonumber\\    
N_d =1&(0, \ \  \epsilon)(2, 3)(2, \ \ -1)  
\label{consist}
\eeqa
\end{center} 
However, as we will argue later the choise (\ref{no2}), or (\ref{no3})
is more natural, as the choise (\ref{no1}) gives that the number of
electroweak Higges present in the models is zero, an unnatural choise.

\item {\em Tadpoles for PS-B classes of models}

The solution to the tadpole 
constraints depend 
on four 
integer parameters $n_a^2$, $n_d^2$, $n_b^1$, $n_c^1$, the phase parameter
$\epsilon = \pm 1$, the parameter $\rho =1, 1/3$ and the NS-background
parameter
$\beta =\ 1-b_i$, which is associated to the parametrization of the 
NS B-field by $b_i =0,\ 1/2$, and the condition $\alpha \gamma =\ 4$.
The latter condition effectively gives the set of values
\beq
\alpha \gamma =\ [({\underline{\pm1,\pm 4}}), (\pm2, \pm2)],
\label{condition}  
\eeq
where by underline we denote permutation of entries.

We note that the presented two different classes of solutions to the
tadpoles, are distinguished by the fixed positive or negative entry 
$m$-wrapping in the colour a-brane.

\begin{table}[htb]\footnotesize
\renewcommand{\arraystretch}{2}
\begin{center}
\begin{tabular}{||c||c|c|c||}
\hline
\hline
$N_i$ & $(n_i^1, m_i^1)$ & $(n_i^2, m_i^2)$ & $(n_i^3, m_i^3)$\\
\hline\hline
 $N_a=4$ & $(0, \epsilon)$  &
$(n_a^2, 3 \epsilon \b_2)$ & $({\tilde \epsilon}, {\tilde \epsilon}/2)$  \\
\hline
$N_b=2$  & $(-1, \epsilon m_b^1 )$ & $(1/\beta_2, 0)$ &
$({\tilde \epsilon}, {\tilde \epsilon}/2)$ \\
\hline
$N_c=2$ & $(1, \epsilon m_c^1 )$ &   $(1/\beta_2, 0)$  & 
$({\tilde \epsilon}, -{\tilde \epsilon}/2)$ \\    
\hline
$N_d=1$ & $(0, \epsilon)$ &  $(n_d^2, 6 \epsilon \b_2)$  
  & $(-2{\tilde \epsilon}, {\tilde \epsilon})$  \\   
\hline
\end{tabular}
\end{center}
\caption{\small
Tadpole solutions for PS-A type models with
D6-branes wrapping numbers giving rise to the 
fermionic spectrum of the type I model, with the SM,
$SU(3)_C \times SU(2)_L \times U(1)_Y$, gauge group at low energies.
The wrappings 
depend on two integer parameters, 
$n_a^2$, $n_d^2$, the NS-background $\beta_i$ and the 
phase parameters $\epsilon = {\tilde \epsilon }= \pm 1$. 
Also there is an additional dependence on the two wrapping
numbers, integer of half integer,
$m_b^1$, $m_c^1$.
\label{spectruma101}}
\end{table}

\begin{table}[htb]\footnotesize
\renewcommand{\arraystretch}{3}
\begin{center}
\begin{tabular}{||c||c|c|c||}
\hline
\hline
$N_i$ & $(n_i^1, m_i^1)$ & $(n_i^2, m_i^2)$ & $(n_i^3, m_i^3)$\\
\hline\hline
 $N_a=4$ & $(1/\beta_1, 0)$  &
$(n_a^2,  -\epsilon \b_2)$ & $(1/\rho, \frac{3{\tilde \epsilon}\rho}{{2}})$  \\
\hline
$N_b=2$  & $(n_b^1, \epsilon \b_1)$ & $(1/\beta_2, 0)$ &
$({\tilde \epsilon}/\rho, \frac{3\rho}{{2}})$ \\
\hline
$N_c=2$ & $(n_c^1, \epsilon \b_1)$ &   $(1/\beta_2, 0)$  & 
$({\tilde \epsilon}/\rho, -\frac{3\rho}{{2}})$ \\    
\hline
$N_d=1$ & $(\alpha/{\beta_1}, 0)$ &  $(n_d^2,  -\gamma \epsilon \b_2)$  
  & $(1/{\rho}, -\frac{3{\tilde \epsilon}\rho}{{2}})$  \\   
\hline
\end{tabular}
\end{center}
\caption{\small Tadpole solutions of 
PS-B type models with D6-branes wrapping numbers giving rise to the 
fermionic spectrum of type I model, with an
$SU(3)_C \times SU(2)_L \times U(1)_Y \times U(1)$ gauge group
at low energies, the extra $U(1)$ being anomaly free.
The parameter $\rho$ takes the values $1, 1/3$, while there 
is an additional dependence on four integer parameters, 
$n_a^2$, $n_d^2$, $n_b^1$, $n_c^1$, the NS-background $\beta_i$, $i=1,2$,
 and
the phase parameters $\epsilon = \pm 1$, ${\tilde \epsilon} = \pm 1$.
Note the condition $\alpha \gamma=4$
and the positive wrapping number entry on the 3rd tori of 
the colour a-brane.  
\label{spectrum10}}
\end{table}

In the rest of this section we will be examining the tadpole 
solutions of the models described in table (\ref{spectrum10}).
The choises of wrapping numbers of table (\ref{spectrum10}) satisfy all the 
tadpole constraints.
The first tadpole condition
in (\ref{na1}) reads \footnote{We have added an arbitrary number
of $N_D$ branes which do not contribute to the rest of the tadpoles and
intersection numbers. This is always an allowed choise. We chosen not
to exhibit the rest of the tadpoles as they involve the identity $0=0$.
Also we have chosen ${\tilde \epsilon} =1$.}
\beq
\frac{4 n_a^2}{\rho \b_1} + 2 \frac{n_b^1}{\rho \b_2} +2 
\frac{n_c^1}{\rho \b_2} + \frac{\alpha n_d^2}{\rho \b_1}+ N_D n_1 n_2 n_3=16.
\label{ena11}
\eeq
To see clearly the cancellation of tadpoles, we have to choose a
consistent numerical set of wrapping
numbers, e.g
\beq
\rho =\ \epsilon =\ 1,\;n_a^2=0,\;n_b^1=0,\;n_c^1=1,\;n_d^2=-1,\;\b_2=1,
\;\b_1=1,\; \alpha =2,\;\gamma=2.
\label{numero1}
\eeq
With the above choise, all tadpole conditions are satisfied but the first, 
which gives
\beq
N_D n_1 n_2 n_3 = 16,
\eeq
The latter can be satisfied with the addition of eight D6-branes
with wrapping numbers $(1, 0)(1, 0)(2, 0)$, effectively giving to the 
models
the structure 
\begin{center}
\beqa
N_a =4&(1, 0)(0, -1)(1,  3/2) \nonumber\\
N_b =2&(0, \ \ 1)(1, \ 0) (1, \ 3/2) \nonumber\\
N_c =2&(1,\ 1)(1,\ 0)(1, -3/2) \nonumber\\    
N_d =1&(2, \ 0)(-1, -2)(1, -3/2)\nonumber\\
N_D =8&(1,\ \  0) (1,\  \ 0)(2, \ \  0)  
\label{consist234}
\eeqa
\end{center} 
Alternatively, we can choose
\beq
\rho =\ \epsilon =\ 1,\;n_a^2=0,\;n_b^1=0,\;n_c^1=1,\;n_d^2=1,\;\b_2=1,
\;\b_1=1,\; \alpha =2,\;\gamma=2.
\label{numero2}
\eeq
With the above choise, all tadpole conditions are satisfied but the first, 
which gives
\beq
4 + N_D n_1 n_2 n_3 = 16,
\eeq
The latter can be satisfied with the addition of six D6-branes
with wrapping numbers $(1, 0)(1, 0)(2, 0)$, effectively giving the model
the structure 
\begin{center}
\beqa
N_a =4&(1, 0)(0, -1)(1,  3/2) \nonumber\\
N_b =2&(0, \ \ 1)(1, \ 0) (1, \ 3/2) \nonumber\\
N_c =2&(1,\ 1)(1,\ 0)(1, -3/2) \nonumber\\    
N_d =1&(2, \ 0)(1, -2)(1, -3/2)\nonumber\\
N_D =6&(1,\ \  0) (1,\ \ 0)(2, \ \  0)  
\label{consist1}
\eeqa 
\end{center}

Note that it appears that 
the wrapping number $(2, 0)$ along the first tori gives rise to 
an additional $U(1)$
at low energies. However, as we will explain in the next section, 
this is an artifact of 
the procedure as its presence can be absorbed into the surviving, the GS
mechanism, massless anomaly free 
$U(1)$ 
field, by a proper field redefinition.

\end{itemize}

f) the hypercharge operator for PS-A, PS-B classes of models
 is defined as a linear combination
of the three diagonal generators of the $SU(4)$, $SU(2)_L$, $SU(2)_R$ groups:
\beq
Y = \frac{1}{2}T_{3R}+ \frac{1}{2}T_{B-L},\;T_{3R}=diag(1,-1),\;
T_{B-L}=diag(\frac{1}{3},\frac{1}{3},\frac{1}{3}, -1). 
\label{hyper12}
\eeq 
Explicitly,
\beqa
Q & = & Y   +\ \frac{1}{2}T_{3L}.\\
\label{hye1}
\eeqa

\section{Cancellation of U(1) Anomalies}

The mixed anomalies $A_{ij}$ of the four $U(1)$'s
with the non-Abelian gauge groups are given by
\beq
A_{ij}= \frac{1}{2}(I_{ij} - I_{i{j^{\star}}})N_i.
\label{ena9}
\eeq
Moreover, analyzing the mixed anomalies 
of the extra $U(1)$'s with the non-abelian gauge groups $SU(4)_c$, 
$SU(2)_R$, $SU(2)_L$ we can see that there are two anomaly free combinations
$Q_b - Q_c$, $Q_a - Q_d$.
Note that gravitational anomalies cancel since D6-branes never 
intersect O6-planes.
In the orientifolded type I torus models gauge anomaly 
cancellation \cite{iru} proceeds through a 
generalized GS
mechanism \cite{louis2} that makes use of the 10-dimensional RR gauge fields
$C_2$ and $C_6$ and gives at four dimensions
the couplings to gauge fields 
 \beqa
N_a m_a^1 m_a^2 m_a^3 \int_{M_4} B_2^o \wedge F_a &;& n_b^1 n_b^2 n_b^3
 \int_{M_4}
C^o \wedge F_b\wedge F_b,\\
N_a  n^J n^K m^I \int_{M_4}B_2^I\wedge F_a&;&n_b^I m_b^J m_b^K \int_{M_4}
C^I \wedge F_b\wedge F_b\;,
\label{ena66}
\eeqa
where
$C_2\equiv B_2^o$ and $B_2^I \equiv \int_{(T^2)^J \times (T^2)^K} C_6 $
with $I=1, 2, 3$ and $I \neq J \neq  K $. Notice the four dimensional duals
of $B_2^o,\ B_2^I$ :
\beqa
C^o \equiv \int_{(T^2)^1 \times (T^2)^2 \times (T^2)^3} C_6&;C^I \equiv
\int_{(T^2)^I} C_2, 
\label{ena7}
\eeqa
where $dC^o =-{\star} dB_2^o,\; dC^I=-{\star} dB_2^I$.

The triangle anomalies (\ref{ena9}) cancel from the existence of the
string amplitude involved in the GS mechanism \cite{sa} in four 
dimensions \cite{iru}. 
The latter amplitude, where the $U(1)_a$ gauge field couples to one
of the propagating
$B_2$ fields, coupled to dual scalars, that couple in turn to
two $SU(N)$ gauge bosons, is 
proportional \cite{louis2} to
\beq
-N_a  m^1_a m^2_a m^3_a n^1_b n^2_b n^3_b -
N_a \sum_I n^I_a n^J_a n^K_b m^I_a m^J_b m^K_b\; ,
I \neq J, K 
\label{ena8}
\eeq

The study of $U(1)$ anomalies in the models is performed separately
for PS-A, PS-B models. We distinguish two cases :

$\bullet$ {\em PS-A models}

For this class of models
the RR couplings $B_2^I$ of (\ref{ena66}), appear 
into three terms (we set for simplicity ${\tilde \epsilon}=1$)  :
\beqa
B_2^3 \wedge \left( \frac{ \epsilon {\tilde \epsilon}}{\b_2 } 
\right)(F^b + F^c),&\nonumber\\
B_2^1 \wedge \left(\epsilon {\tilde \epsilon}[ 
4 n_a^2 \ F^a - 2n_d^2 \ F^d +
\frac{2 m_b^1}{\b_2} \ F^b + \frac{2 m_c^1}{\b_2}\ F^c]
\right),&\nonumber\\
B_2^o  \wedge \left(6\b_2 {\tilde \epsilon} \right) (F^a + F^d).&
\label{rr23}
\eeqa

As can be seen from (\ref{rr23}) two anomalous combinations of $U(1)$'s,
e.g.
 $F^a + F^d$, $F^b + F^c$ become massive through their couplings to RR
 fields $B_2^o$, $B_2^3$. Also there are
two non-anomalous $U(1)$'s, the combinations of
$Q^b - Q^c$, $Q^a - Q^d$. A third non-anomalous combination of
$U(1)$'s is made massive by its coupling to
$B_2^1$.

At this point we should list 
the couplings of the dual scalars $C^I$ of $B_2^I$ required to cancel
the mixed anomalies of the four $U(1)$'s with the 
non-abelian gauge groups $SU(N_a)$. They are given by

\beqa
C^o \wedge \left( \frac{ {\tilde \epsilon}}{ \b_2}\right) [-(F^b \wedge 
F^b) + (F^c \wedge F^c)], &\nonumber\\
C^3 \wedge  \left(3 {\tilde \epsilon} \b_2\right) [ (F^a \wedge 
F^a)-4 (F^d \wedge F^d)],
&\nonumber\\
C^2 \wedge [  \frac{n_a^2{\tilde \epsilon}{\epsilon} }{2}(F^a \wedge 
F^a)  +     \frac{ m_b^1{\tilde \epsilon} {\epsilon} }{2 \b_2}(F^b \wedge 
F^b)  - \frac{ m_c^1  {\tilde \epsilon} {\epsilon}}{2 \b_2}(F^c \wedge 
F^c)  - 2n_a^2{\tilde \epsilon} {\epsilon} (F^d \wedge F^d) ].&
\label{nonanomal}
\eeqa

Note that the combination of $U(1)$'s which survives
massless to low energies
is uniquely given by
\beq
Q_l =\ \kappa \ ( (Q_a -\ Q_d) +\  (Q_b -\ Q_c) ),
\label{survi1}
\eeq
where $\kappa$ an arbitrary number.

$\bullet$ {\em PS-B models}

If we take into account the phenomenological requirements of eqn.
(\ref{req1})  the RR couplings $B_2^I$ of (\ref{ena66}), appear 
into three terms \footnote{We set for simplicity ${\tilde \epsilon} =1$.} :
\beqa
B_2^1 \wedge \left( \frac{2 \epsilon \b_1 }{\b_2 \rho} 
\right)(F^b + F^c),&\nonumber\\
B_2^2 \wedge \left( \frac{-4 \epsilon \b_2}{\b_1 \rho}  
\right)(F^a + F^d),&\nonumber\\
B_2^3  \wedge \left( \frac{3 \rho}{\b_2}  \right)\left(    
\frac{2 n_a^2 \b_2 F^a}{\b_1} + n_b^1 F^b - n_c^1 F^c -
\frac{\b_2 \alpha n_d^2}{2 \b_1} F^d
  \right).&
\label{rr1}
\eeqa
At this point we should list 
the couplings of the dual scalars $C^I$ of $B_2^I$ required to cancel
the mixed anomalies of the four $U(1)$'s with the 
non-abelian gauge groups $SU(N_a)$. They are given by
\beqa
C^1 \wedge [ \left( \frac{-3 \epsilon \b_2 \rho}{2 \b_1} \right)(F^a \wedge 
F^a - 4 F^d \wedge F^d)], &\nonumber\\
C^2 \wedge \left( \frac{3 \b^1 \rho \epsilon}{2 \b_2 }  \right)[(F^b \wedge 
F^b)
-(F^c \wedge F^c)],
&\nonumber\\
C^o \wedge \left(  \frac{n_a^2}{\rho \b_1}(F^a \wedge 
F^a)  +     \frac{n_b^1}{\rho \b^2}(F^b \wedge 
F^b)  + \frac{n_c^1}{\rho \b_2}(F^c \wedge 
F^c)  + \frac{\alpha n_d^2}{\rho \b_1}(F^d \wedge F^d) \right),&
\label{rr2}
\eeqa

Notice that the RR scalar $B_2^0$ does not couple to any 
field $F^i$ as we have imposed the condition (\ref{req1}) 
which prevents the appearance of any exotic 
matter.\newline
Looking at (\ref{rr1}) we conclude that
there are two anomalous $U(1)$'s, $Q^b + Q^c$, $Q^a + Q^d$,
which become massive through their couplings to the 
RR 2-form fields $B_2^1, B_2^2$
and two non-anomalous free combinations
$Q^b - Q^c$, $Q^a - Q^d$.
Note that the 
 mixed anomalies $A_{ij}$ are cancelled 
by the GS mechanism set by the couplings (\ref{rr1}, \ref{rr2}).
In addition, the combination of the $U(1)$'s which remains light at low 
energies, and
is orthogonal to the massive $U(1)$'s coupled to the RR fields $B_2^3$, 
$B_2^2$, $B_2^1$
is 
\beq
n_b^1+n_c^1 \neq 0,\;\ Q_l = \frac{1}{(n_b^1+ n_c^1) }(Q_b - Q_c)
-\frac{ \b_1}{\b_2(2 n_a^2 + \frac{\alpha n_d^2}{2})}(Q_a -Q_d).
\label{hyper}
\eeq
Making the choise of wrapping numbers (\ref{numero1}),
the surviving massless non-anomalous $U(1)$ reads
\beq
Q_l =\ (Q_b -\ Q_c) +\ (Q_a -\ Q_d).
\label{asda12}
\eeq
Instead, if we make the choise (\ref{numero2}) 
the surviving massless non-anomalous $U(1)$ reads
\beq
Q_l =\ (Q_b -\ Q_c) -\ (Q_a -\ Q_d).
\label{asda14}
\eeq
Both choises of global $U(1)$'s are consistent with electroweak data
in the sense that they don't break the lepton number. That happens because 
the 
bidoublet Higgs fields $h_1$, $h_2$ don't get charged. A similar effect holds
for (\ref{survi1}).

A comment is in order. The interpretation of the presence of the 
$(2, 0)$ wrapping number, in (\ref{consist234}), (\ref{consist1})
found in the
d-brane of 
the first tori \footnote{Note that there is no NSNS b-field in the 
first torus.} is subttle in principle. That happens since
D-brane gauge theory analysis indicates that it should be interpreted either 
as one brane wrapping twice around the cycle $(1, 0)$ or as two branes 
wrapping
once around the cycle $(1, 0)$ giving rise to two $U(1)$'s, $Q_d^1$, $Q_d^2$, 
where the two $U(1)$'s correspond to the combinations
\beq
Q_d^{(1)} = (Q_d^{1}+ Q_d^{2});\;\;Q_d^{(2)}=(Q_d^{1}- Q_d^{2}).
\label{twoone}
\eeq 
The two $U(1)$'s listed in (\ref{twoone}) correspond
to open strings stretching between the first wrapping of the d-brane, namely 
$Q_d^{(1)}$, and the first wrapping of the extra brane, namely $Q_d^{(2)}$.

However, for the string GUT model which starts at the string scale 
with four $U(1)$'s, it is only
tadpole cancellation that introduces an additional $U(1)$, from 
``multiwrapping''. The additional $U(1)$ was not needed at the gauge theory 
level, as cancellation of the mixed $U(1)$ gauge 
anomalies was already consistent without the need of adding an extra $U(1)$.
Clearly, at the level of the effective action we shouldn't 
have found any additional $U(1)$'s beyond those, four, already present at the 
string scale \footnote{Note that there was no stringy Higgs effect present
that could introduce additional gauge bosons.}.
 
Lets us now redefine the 
massless non-anomalous $U(1)$ as
\beqa
Q_l \rightarrow Q_l =\ (Q_b -\ Q_c) +\ 
\left(Q_a -\frac{1}{2}(Q_d^{(1)}+Q_d^{(2)})\right),\\
Q_l \rightarrow Q_l =\ (Q_b -\ Q_c) +\ \left(Q_a -\frac{1}{2}( Q_d^1 + 
Q_d^2 + Q_d^1 -Q_d^2)\right), 
\eeqa
where is is clear that we have identify $Q_d = Q_d^1$.
Let us 
rewrite the charges of the fermion fields
of table one, as 
\beqa
(4, 2, 1)_{[1, 1, 0, {\underline{0, 0}}   ]  }, &
({\bar 4}, 1, 2)_{[-1, 0, 1, {\underline{0, 0}}]},\nonumber\\
(1, {\bar 2}, 1)_{[0, -1, 0, {\underline{1, 0}}]},&
(1, 1, {\bar 2})_{[0, 0, -1, {\underline{-1, 0}}[},
\label{unde}
\eeqa
where by underline we indicate a simultaneous permutation of the fourth, 
fifth entries for all fermion fields. Thus no additional charges are \
introduced for the fields
beyond the already present.    
 It is now clearly 
seen that
the additional $U(1)$, from ``multiwrapping'' corresponds just
to a field redefinition of the surviving global $U(1)$ at low 
energies and hence at 
the level of the effective action at low energy has no physical effect. 
In fact, at the level the cancellation of the mixed global $U(1)$ 
gauge anomalies its time either $Q_d^{(1)}$ 
or $Q_d^{(2)}$ get charged.

Let us close this section by noticing that the non-anomalous 
massless $U(1)$ which is free from gauge 
and gravitational anomalies can be written in three more different ways. 
We enumerate them here for consistency. They read :
\beqa
\b_1 \neq 0,& Q_l =\ \frac{1 }{ \b_1 }(Q_b - Q_c) - \frac{n_c^1+ n_b^1}
{ \b_2(2 n_a^2  + \frac{\alpha n_d^2}{2})  }(Q_a - Q_d),\\
\b^2(2n_a^2 + \frac{\alpha n_d^2}{2}) \neq 0,& Q_l = \b^2(2n_a^2 
+ \frac{n_d^2}{2})
(Q_b -Q_c)-\frac{(n_c^2 + n_b^1)}{(\b_1)^{-1}}(Q_a - Q_d),\\
\frac{\b_2(2 n_a^2  +  \frac{\alpha n_d^2}{2}) }{\b_1} \neq 0,&
Q_l =\ \frac{\b_2(2 n_a^2  +  \frac{\alpha n_d^2}{2})}{(\b_1)}
(Q_b -Q_c) -  ( n_b^1  + n_c^1)(Q_a - Q_d).
\label{hyper2}  
\eeqa

\section{Higgs sector, global symmetries, proton stability, $N=1$ SUSY on
intersections and
neutrino masses}

\subsection{Stability of the configurations and Higgs sector}

We have so far seen the appearance in the R-sector 
of $I_{ab}$ massless fermions
in the D-brane intersections 
transforming under bifundamental representations $N_a, {\bar N}_b$.
 In intersecting 
brane words, besides the actual
presence of massless fermions at each intersection, 
we have evident the presence of an equal number of
 massive bosons, in the NS-sector, in the same representations 
as the massless fermions \cite{luis1}.
Their mass is of order of the string scale and it should be taken 
into account when examining phenomenological applications related to the
renormalization group equations.
However, it is possible that some of those 
massive bosons may become 
tachyonic \footnote{For consequences
when these set of fields may become massless see \cite{cim}.}, 
especially when their mass, that depends on the 
angles between the branes,
is such that is decreases the world volume of the 
3-cycles involved in the recombination process of joining the two
branes into a single one \cite{senn}.
Denoting the twist vector by $(\vartheta_1,\vartheta_2,
\vartheta_3,0)$, in the NS open string sector the 
lowest lying states are given by \footnote{
we assume $0\leq\vartheta_i\leq 1$ .}
{\small \beqa
\begin{array}{cc}
{\rm \bf State} \quad & \quad {\bf Mass} \\
(-1+\vartheta_1,\vartheta_2,\vartheta_3,0) & \alpha' M^2 =
\frac 12(-\vartheta_1+\vartheta_2+\vartheta_3) \\
(\vartheta_1,-1+\vartheta_2,\vartheta_3,0) & \alpha' M^2 =
\frac 12(\vartheta_1-\vartheta_2+\vartheta_3) \\
(\vartheta_1,\vartheta_2,-1+\vartheta_3,0) & \alpha' M^2 =
\frac 12(\vartheta_1+\vartheta_2-\vartheta_3) \\
(-1+\vartheta_1,-1+\vartheta_2,-1+\vartheta_3,0) & \alpha' M^2
= 1-\frac 12(\vartheta_1+\vartheta_2+\vartheta_3)
\label{tachdsix}
\end{array}
\eeqa}
Exactly at the point, where one of these masses may become massless we have 
pseservation of ${\cal N}=1$ locally. 
The angles at the four different intersections can be expressed
in terms of the parameters of the tadpole solutions.

We note that in the study of Higgs sector, we will deal separately with the definition
of the angle structure for the PS-A, PS-B types of PS models.
However, where it applies we will list the similarities. 

$\bullet$ {\em Angle structure and Higgs fields for PS-A classes of models}

The angles at the different intersections can be expressed in terms of the 
tadpole solution parameters. 
We define the angles:
\beqa
\theta_1 \   = \ \frac{1}{\pi} cot^{-1}\frac{ R_1^{(1)}}{ m_b^1 R_2^{(1)}} \ ;\
\theta_2 \  =   \  \frac{1}{\pi} cot^{-1}
\frac{n_a^2 R_1^{(2)}}{3\beta_2 R_2^{(2)}} \ ;\
\theta_3 \  = \  \frac{1}{\pi} cot^{-1}\frac{2R_1^{(3)}}{  R_2^{(3)}},
 \nonumber \\
{\tilde {\theta_2}} \   = 
\ \frac{1}{\pi} cot^{-1}\frac{n_d^2 R_1^{(1)}}{ 3 \b_2 R_2^{(1)}}
\ ;\
{\tilde {\theta_1}} \   = 
\ \frac{1}{\pi} cot^{-1}\frac{ R_1^{(1)}}{ m_c^1 R_2^{(1)}}\ ,\
\label{angPSA}
\eeqa
where $R_{i}^{(j)}$, $i={1,2}$ are the compactification radii
for the three $j=1,2,3$ tori, namely
projections 
of the radii 
 onto the cartesian axis $X^{(i)}$ directions when the NS flux B field,
$b^k$, $k=1,2$ is turned on. 

At each of the four non-trivial intersections 
we have the 
presense of four states $t_i , i=1,\cdots, 4$, associated
to the states (\ref{tachdsix}).
 Hence we have a total of
sixteen different scalars in the model.
The setup is seen clearly if we look at figure one.
These scalars are generally massive but for some values of
their angles could become tachyonic (or massless).

Also, if we demand that the scalars associated with (\ref{tachdsix}) and 
PS-A models 
may not be tachyonic,
we obtain a total of twelve 
conditions for the PS-A type models
 with a D6-brane at angles
configuration to be stable.
They are
given in Appendix I.
We don't consider
the scalars from 
the $aa^{\star}$, $dd^{\star}$ intersections. For these sectors
we will require later that they preserve $N=1$ SUSY. 
As a result all scalars in these sectors may become massive for both
PS-A, PS-B models.

%%%%%%%%%% Figure here%%%%%%%%%%%%%%%%%%%%%%%%%%%%%%%%%%%%%%%%%
\begin{figure}
\centering
\epsfxsize=6in
\hspace*{0in}\vspace*{.2in}
\epsffile{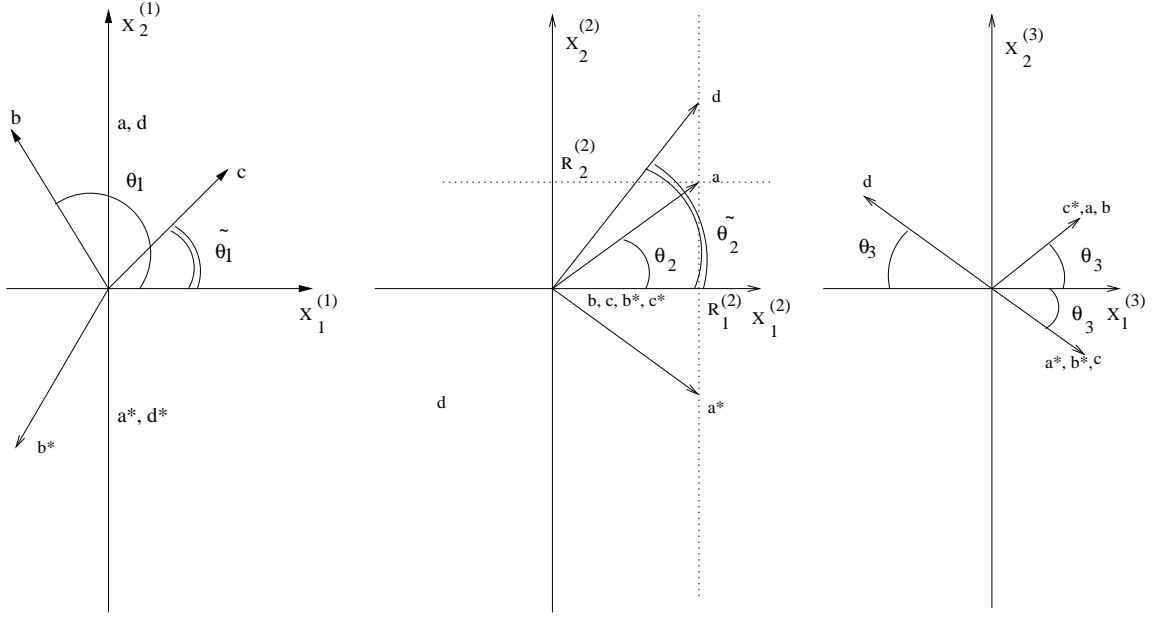}
\caption{\small
Assignment of angles between D6-branes on a
a type I PS-A class of models
based on the initial gauge group $U(4)_C\times {U(2)}_L\times
{U(2)}_R$. The angles between branes are shown on a product of 
$T^2 \times T^2 \times T^2$. We have chosen $ \b_1 =1$, $m_b^1, 
m_c^1, n_a^2 >0$, $\epsilon = {\tilde \epsilon}= 1$. These models
break to low energies to exactly the SM.}
\end{figure}
%%%%%%%%%%%%%%%%%end of figure %%%%%%%%%%%%%%%%%%%%%%%%%%%%%%%%%%%

$\bullet$ {\em Angle structure and Higgs fields for PS-B classes of models}

Let us define the angles :
\beqa
\theta_1 \   = \ \frac{1}{\pi} cot^{-1}
\frac{n_b^1R_1^{(1)}}{\beta^1R_2^{(1)}} \ ;\
\theta_2 \  =   \  \frac{1}{\pi} cot^{-1}
\frac{n_a^2R_1^{(2)}}{\beta_2 R_2^{(2)}} \ ;\
\theta_3 \  = \  \frac{1}{\pi} cot^{-1}\frac{2R_1^{(3)}}{3 \rho R_2^{(3)}},
 \nonumber \\
{\tilde {\theta_1}} \   = 
\ \frac{1}{\pi} cot^{-1}\frac{n_c^1 R_1^{(1)}}{ \beta_1 R_2^{(1)}}\ ;\
{\tilde { \theta_2}} \ 
 = \ \frac{1}{\pi} cot^{-1}\frac{n_d^2 R_1^{(2)}}{4 \beta_2 R_2^{(2)}}   \ ;\
{\tilde {\theta_3 }} \  = \ \frac{1}{\pi} cot^{-1}
\frac{2R_1^{(3)}}{3\rho R_2^{(3)}},
\label{angulos}
\eeqa
where $R^{(i)}_{1,2}$ are the compactification radii
for the three $i=1,2,3$ tori, namely
projections 
of the radii 
 onto the $X^{(i)}_{1,2}$ directions when the NS flux B field,
$b^i$, is turned on.

At each of the four non-trivial intersections 
we have the 
presense of four states $t_i , i=1,\cdots, 4$, associated
to the states (\ref{tachdsix}).
 Hence we have a total of
sixteen different scalars in the model.
The setup is seen clearly if we look at figure two.

In addition, some interesting relations between the different
scalar fields hold e.g for PS-B models:
\beqa
m^2_{c d^{\star}}(t_2) + m^2_{cd^{\star}}(t_3) = m^2_{a c}(t_2) +
m^2_{ac }(t_3)      &\nonumber\\
m^2_{a b^{\star}}(t_1) + 
m^2_{a b^{\star}}(t_3)=  m^2_{ac}(t_1) + m^2_{a c}(t_3) &\nonumber\\
m^2_{c d^{\star}}(t_2) + m^2_{b d}(t_3) =   m^2_{c d^{\star}}(t_3)+
m^2_{b d}(t_2)&\nonumber\\
m^2_{a b^{\star}}(t_2) + m^2_{a b{\star}}(t_3) =   m^2_{bd}(t_2)+
m^2_{b d}(t_3)&\nonumber\\
\label{srela}
\eeqa

%%%%%%%%%% Figure here%%%%%%%%%%%%%%%%%%%%%%%%%%%%%%%%%%%%%%%%%
\begin{figure}
\centering
\epsfxsize=6in
\hspace*{0in}\vspace*{.2in}
\epsffile{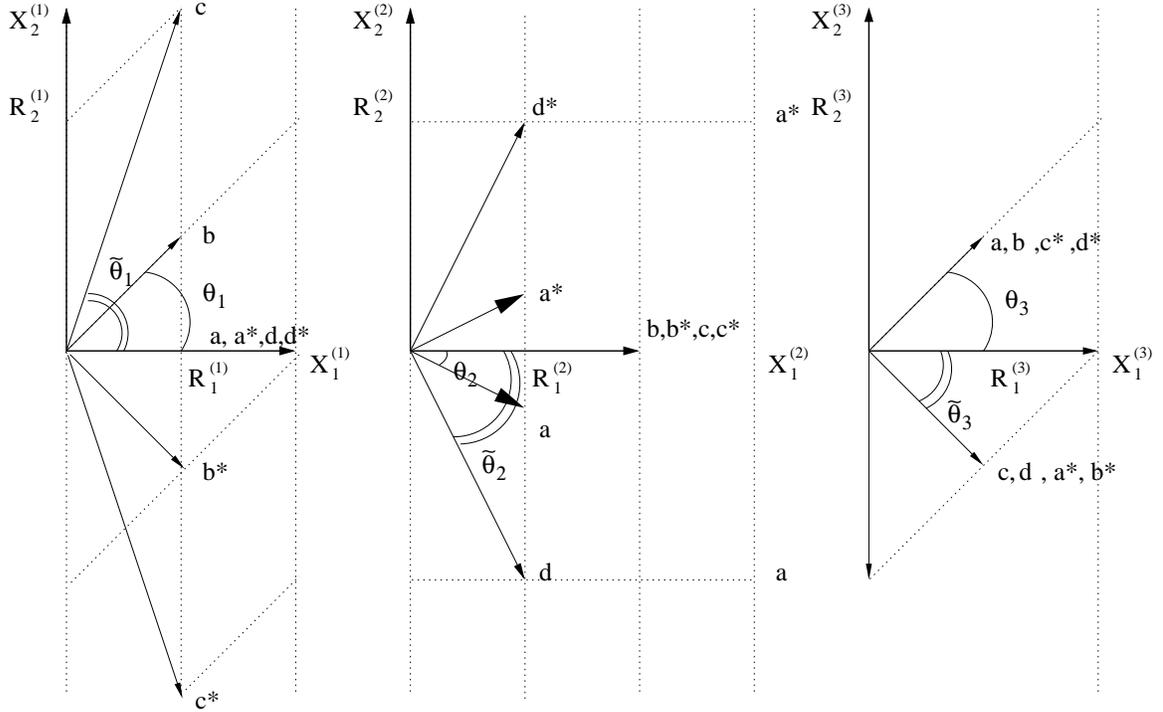}
\caption{\small
Assignment of angles between D6-branes on a
a type I PS-B class of models
based on the initial gauge group $U(4)_C\times {U(2)}_L\times
{U(2)}_R$. The angles between branes are shown on a product of 
$T^2 \times T^2 \times T^2$. We have chosen $\rho = \b_1 =1$, $n_b^1, 
n_c^1, n_a^2, 
n_d^2 >0$, $\epsilon = 1$. These models break to low energy
to the SM augmented by an anomaly free $U(1)$ symmetry.
}
\end{figure}
%%%%%%%%%%%%%%%%%end of figure %%%%%%%%%%%%%%%%%%%%%%%%%%%%%%%%%%%

Demanding that the scalars associated with (\ref{tachdsix}) in PS-B models 
may not be tachyonic,
we obtain a total of twelve 
conditions for a D6-brane at angles configuration to be stable.
They are
given in Appendix II.

Lets us now turn our discussion to the Higgs sector of 
PS-A, PS-B models.
In general there are two different Higgs fields that may
be used to break the PS symmetry.
We remind that they were given in (\ref{useful1}).
The question is if $H_1$, $H_2$ are present in the spectrum of 
PS-A, PS-B models. The following discussion unless otherwise stated it will 
apply for both classes of models.
In general, tachyonic scalars stretching between two 
different branes $\tilde a$, 
$\tilde b$, can be used as Higgs scalars as they can become non-tachyonic
by varying the distance between the branes.
Looking at the $I_{a c^{\star}}$ intersection we can answer positively
to our question since there are scalar doublets $H^{\pm}$ localized.
They come from open strings
stretching
between the $U(4)$ $a$-brane and $U(2)_R$ $c^{\star}$-brane.

\begin{table} [htb] \footnotesize
\renewcommand{\arraystretch}{1}
\begin{center}
\begin{tabular}{||c|c||c|c|c|c||}
\hline
\hline
Intersection & PS breaking Higgs & $Q_a$ & $Q_b$ & $Q_c$ & $Q_d$ \\
\hline\hline
$a c^{\star}$ & $H_1$  &
$1$ & $0$ & $1$ & $0$ \\
\hline
$a c^{\star}$  & $H_2$  & $-1$ & $0$ & $-1$ & $0$  \\
\hline
\hline
\end{tabular}
\end{center}
\caption{\small Higgs fields responsible for the breaking of 
$SU(4) \times SU(2)_R$ 
symmetry of the 
$SU(4)_C \times SU(2)_L \times SU(2)_R$ type I model with D6-branes
intersecting at angles. These Higgs are responsible for giving
masses to the right handed
neutrinos in a single family.
\label{Higgs}}
\end{table}

The $H^{\pm}$'s come from the NS sector and
correspond to the states \footnote{a similar set of states was used
in \cite{louis2} to provide the model with electroweak Higgs scalars.}  
{\small \beqa
\begin{array}{cc}
{\rm \bf State} \quad & \quad {\bf Mass^2} \\
(-1+\vartheta_1, \vartheta_2, 0, 0) & \alpha' {\rm (Mass)}^2_{H^{+}} =
  { {Z_3}\over {4\pi ^2}}\ +\ \frac{1}{2}(\vartheta_2 - 
\vartheta_1) \\
(\vartheta_1, -1+ \vartheta_2, , 0, 0) & \alpha' {\rm (Mass)}^2_{H^{-}} =
  { {Z_3}\over {4\pi ^2 }}\ +\ \frac{1}{2}(\vartheta_1 - \vartheta_2) \\
\label{Higgsmasses}
\end{array}
\eeqa}
where $Z_3$ is the distance$^2$ in transverse space along the third torus, 
$\vartheta_1$, $\vartheta_2$ are the (relative)angles 
between the $a$-, $c^{\star}$-branes in the 
first and second complex planes respectively.  
The presence of scalar doublets $H^{\pm}$ can be seen as
coming from the field theory mass matrix

\beq
(H_1^* \ H_2) 
\left(
\bf {M^2}
\right)
\left(
\begin{array}{c}
H_1 \\ H_2^*
\end{array}
\right) + h.c.
\eeq
where
\beqa
{\bf M^2}=M_s^2
\left(
\begin{array}{cc}
Z_3^{(ac^*)}(4 \pi^2)^{-1}&
\frac{1}{2}|\vartheta_1^{(ac^*)}-\vartheta_3^{(ac^*)}|  \\
\frac{1}{2}|\vartheta_1^{(ac^*)}-\vartheta_3^{(ac^*)}| &
Z_3^{(ac^*)}(4 \pi^2)^{-1}\\
\end{array}
\right),
\eeqa
\vspace{1cm}
The fields $H_1$ and $H_2$ are thus defined as
\beq
H^{\pm}={1\over2}(H_1^*\pm H_2) 
\eeq
where their charges are given in table (\ref{Higgs}). 
Hence the effective potential which 
corresponds to the spectrum of the PS symmetry breaking
Higgs scalars is given by
\beqa
V_{Higgs}\ =\ m_H^2 (|H_1|^2+|H_2|^2)\ +\ (m_B^2 H_1H_2\ +\ h.c)
\label{Higgspot}
\eeqa
where
\beqa 
{m_H}^2 \ =\ {{Z_3^{(ac^*)}}\over {4\pi ^2\alpha '}} \ & ;&
m_B^2\ =\ \frac{1}{2\alpha '}|\vartheta_1^{(ac^*)}-\vartheta_2^{(ac^*)}|
\label{masillas}
\eeqa
The precise values of $m_H^2$, $m_B^2$, for PS-A models, are :

\beqa
 {m_H}^2 \ \stackrel{PS-A}{=}\ {
 {(\xi_a^{\prime}+\xi_c^{\prime})^2}\over {\alpha '}}\ ;\
m_B^2\ \stackrel{PS-A}{=}\ \frac{1}{2\alpha '}|\frac{1}{2}+ {\tilde\theta}_1 - {\theta}_2|\ ,
\label{value100}
\eeqa
where $\xi_a^{\prime}$($\xi_c^{\prime}$) is the distance between the
orientifold plane
and the $a$($c$) branes and ${\tilde \theta}_1$, 
$ {\theta}_2$ were defined in
(\ref{angPSA}). In terms of those data for PS-A models we found :

\beqa
m_B^2 &\stackrel{PS-A}
{=}&
\frac{1}{2}|
m^2_{\chi_R} (t_2) +\ m^2_{\chi_R}(t_3)
-\ m^2_{{\bar F}_R}(t_1) -\
m^2_{{\bar F}_R}(t_3)|\nonumber\\
&=&
\frac{1}{2}|
m^2_{\chi_R} (t_2) +\ m^2_{\chi_R}(t_3)
-\ m^2_{{F}_L}(t_1) -\
m^2_{{F}_L}(t_3)|\nonumber\\
\label{kainour}
\eeqa

For PS-B models, 
\beqa
 {m_H}^2 \ \stackrel{PS-B}{=}\ { {(\xi_a+\xi_c)^2}\over {\alpha '}}\ ;\
m_B^2\ \stackrel{PS-B}{=}\ \frac{1}{2\alpha '}|\tilde\theta_1 - {\theta}_2|\ ,
\label{value1001}
\eeqa
where $\xi_a$($\xi_c$) is the distance between the orientifold plane
and the $a$($c$) branes and ${\tilde \theta}_1$, 
$ {\theta}_2$ were defined in
(\ref{angulos}).

The $m_B^2$ mass can be expressed in terms of the scalar masses
(\ref{tachdsix}) present, using the relations
(\ref{srela}).
Explicitly we found :

\beqa
m_B^2 &\stackrel{PS-A}{=}& \frac{1}{2}|m^2_{F_L}(t_2) -
m^2_{F_L}(t_1)|
\label{forthe}
\eeqa

\beqa
m_B^2 &\stackrel{PS-B}{=}& \frac{1}{2}|m^2_{{\bar F}_R}(t_2) -
m^2_{{\bar F}_R}(t_1)|\nonumber\\
&=& \frac{1}{2}| m^2_{\chi_R }(t_2)
+m^2_{  \chi_R  }(t_3) -  m^2_{F_L }(t_1) -
m^2_{F_L }(t_3)|\nonumber\\
& = & \frac{1}{2}| m^2_{{\chi}_R }(t_2)
   +m^2_{{\chi}_R }(t_3) - m^2_{{\bar F}_R}(t_1)-
   m^2_{{\bar F}_R}(t_3)|\nonumber\\    
& = & \frac{1}{2}|m^2_{{\bar F}_R}(t_2)+m^2_{{\bar F}_R}(t_3)
- m^2_{{F}_L }(t_1)- m^2_{{F}_L }(t_3)|\nonumber\\
\label{analytic}
\eeqa

For PS-A, PS-B models the number of Higgs present
is equal to the 
the intersection number product between the  $a$-, $c^{\star}$- branes
in the
first and second complex planes, namely
\beq
n_{H^{\pm}} \stackrel{PS-A}{=}\  I_{ac^{\star}}\ =\ |3\epsilon^2| =\ 3.
\label{inter1}
\eeq
\beq
n_{H^{\pm}} \stackrel{PS-B}{=}\  I_{ac^{\star}}\ =\ |\epsilon^2| =\ 1.
\label{inter12}
\eeq
A comment is in order.   
For PS-A models the number of PS Higgs is three. That means that we have
three
intersections and to each one we have a Higgs particle which is a linear 
combination of the Higgs $H_1$ and $H_2$.
For PS-B models the number of scalar doublets present 
is one, thus the Higgs responsible for
breaking
the PS symmetry will be a linear combination of the $H_1$, $H_2$.

There are, however, more Higgs present. 
In the $bc^{\star}$ intersection we have present some of the most useful
Higgs fields of the model. They will be used later to give mass to the
quarks and leptons of the model. 
They appear in the representations $(1, 2, 2)$,  $(1, {\bar 2}, {\bar 2})$ and
from now on we will 
we denote them as $h_1$, $h_2$.

\begin{table} 
%[htb]
% \footnotesize
%\renewcommand{\arraystretch}{2.5}
\begin{center}
\begin{tabular}{||c|c||c|c|c|c||}
\hline
\hline
Intersection & Higgs & $Q_a$ & $Q_b$ & $Q_c$ & $Q_d$ \\
\hline\hline
$bc^{\star}$  & $h_1 = (1, 2, 2)$  &
$0$ & $1$ & $1$ & $0$ \\
\hline
$bc^{\star}$ & $h_2 = (1, {\bar 2}, {\bar 2})$  & $0$ & $-1$ & $-1$ & $0$  \\
\hline
\hline
\end{tabular}
\end{center}                 
\caption{\small Higgs fields present in the intersection $bc^{\star }$
of the
$SU(4)_C \times SU(2)_L \times SU(2)_R$ type I model with D6-branes
intersecting at angles. These Higgs give masses to the quarks and
leptons in a single family and are responsible for 
electroweak symmetry breaking.
\label{Higgs3}}
\end{table}

In the NS sector
the lightest scalar states $h^{\pm}$ originate from
open strings stretching between the
$bc^{\star }$ branes

{\small \beqa
\begin{array}{cc}
{\rm \bf State} \quad & \quad {\bf Mass^2} \\
(-1+\vartheta_1, 0, 0, 0) & \alpha' {\rm (Mass)}^2 =
  { {{\tilde Z}_{23}^{bc^{\star}}}\over {4\pi^2}}\ -\ \frac{1}{2}(\vartheta_1)  \\
 (\vartheta_1, -1, 0, 0) &  \alpha' {\rm (Mass)}^2 =
  { {{\tilde Z}_{23}^{bc^{\star}}}\over {4\pi^2}} +\  \frac{1}{2}(\vartheta_1)
\label{Higgsacstar}
\end{array}
\eeqa}
where ${\tilde Z}_{23}^{bc^{\star}}$ is the relative distance in 
transverse space along the second and
third torus from the orientifold plane, 
$\th_1$, is the (relative)angle 
between the $b$-, $c^{\star}$-branes in the 
first complex plane.

The presence of scalar doublets $h^{\pm}$ can be seen as
coming from the field theory mass matrix

\beq
(h_1^* \ h_2) 
\left(
\bf {M^2}
\right)
\left(
\begin{array}{c}
h_1 \\ h_2^*
\end{array}
\right) + h.c.
\eeq
where
\beqa
{\bf M^2}=M_s^2
\left(
\begin{array}{cc}
Z_{23}^{(bc^*)}(4\pi^2)^{-1}&
\frac{1}{2}|\vartheta_1^{(bc^*)}-\vartheta_3^{(bc^*)}|  \\
\frac{1}{2}|\vartheta_1^{(bc^*)}-\vartheta_3^{(bc^*)}| &
Z_{23}^{(bc^*)}(4\pi^2)^{-1}\\
\end{array}
\right),
\eeqa
The fields $h_1$ and $h_2$ are thus defined as
\beq
h^{\pm}={1\over2}(h_1^*\pm h_2) \   \ .
\eeq
The effective potential which 
corresponds to the spectrum of electroweak Higgs $h_1$, $h_2$ is given by
\beqa
V_{Higgs}^{bc^{\star}}\ =\ \overline{m}_H^2 (|h_1|^2+|h_2|^2)\ +\ 
(\overline{m}_B^2 h_1 h_2\ +\ h.c)
\label{bcstarpote}
\eeqa
where
\beqa 
\overline{m}_H^2 \ =\ {{{\tilde Z}_{23}^{(bc^*)}}\over {4\pi^2\alpha'}} \ 
& ;&
\overline{m}_B^2\ =\ \frac{1}{2\alpha'}|
\vartheta_1^{(bc^{\star})}|
\label{bchiggs}
\eeqa
The precise values of for PS-A classes of
models $\overline{m}_H^2$, $\overline{m}_B^2$ are
\beqa 
 {\bar m}_{H}^2 \ \stackrel{PS-A}{=}\ { {({\tilde \chi}_b^{(2)}
 +{\tilde \chi}_{c^{\star}}^{(2)})^2 + ({\tilde \xi}_b^{(3)} +
 {\tilde \xi}_{c^{\star}}^{(3)})^2}\over
{\alpha '}}\ ;\
{\bar m}_{B}^2\ \stackrel{PS-A}{=}\ \frac{1}{2\alpha'}
|\tilde\theta_1 + \theta_1 -1|
\ ;
\label{value1002}
\eeqa
where $\tilde\theta_1$, $\theta_1$ were defined in (\ref{angPSA}).
Also ${\tilde \chi}_b, {\tilde \chi}_{c^{\star}}$ are the 
distances of the $b$, $c^{\star}$ branes from the
orientifold plane in the second tori and ${\tilde \xi}_b$,
${\tilde \xi}_{c^{\star}}$
are the distances of the $b, c^{\star}$ branes from the orientifold plane
in the third tori. Notice that the $b$, $c^{\star}$ branes are paralled along
the second and third tori.

The precise values of for PS-B
models $\overline{m}_H^2$, $\overline{m}_B^2$ are
\beqa 
 {{\bar m}_H}^2 \ \stackrel{PS-B}{=}\ { {(\chi_b^{(2)}+\chi_{c^{\star}}^{(2)})^2 + (\xi_b^{(3)} +
 \xi_{c^{\star}}^{(3)})^2}\over
{\alpha '}}\ ;\
{\bar m}_B^2\ \stackrel{PS-B}{=}\ \frac{1}{2\alpha'}|\tilde\theta_1 + \theta_1|
\ ;
\label{value10031}
\eeqa
where $\chi_b, \chi_{c^{\star}}$ are the 
distances of the $b$, $c^{\star}$ branes from the
orientifold plane in the second tori and $\xi_b$, $\xi_{c^{\star}}$ 
are the distances of the $b, c^{\star}$ branes from the orientifold plane
in the third tori. Notice that the $b$, $c^{\star}$ branes are paralled along
the second and third tori.
The angle ${\tilde \theta}_1$, 
was defined in
(\ref{angulos}) and ${\bar m}_B^2$ can be expressed in terms of the scalar masses
of (\ref{tachdsix}) and (\ref{srela}).
We found
\beqa
{\bar m}_B^2\ \stackrel{PS-B}{=}\
\frac{1}{2}
|m^2_{{\bar F}_R}(t_2) +
m^2_{{\bar F}_R}(t_3)  +  m^2_{\chi_L}(t_2) +
m^2_{\chi_L}(t_3)|&\nonumber\\
=\ \frac{1}{2}
|m^2_{{\bar F}_R}(t_2) +
m^2_{{\bar F}_R}(t_3)  +  m^2_{ F_L}(t_2) +
m^2_{F_L}(t_3)|&\nonumber\\
=\ \frac{1}{2}
|m^2_{F_L}(t_2) +
m^2_{F_L}(t_3)  +  m^2_{\chi_R}(t_2) +
m^2_{\chi_R}(t_3)|&\nonumber\\
=\ \frac{1}{2}
|m^2_{\chi_L}(t_2) +
m^2_{\chi_L}(t_3)  +  m^2_{ \chi_R}(t_2) +
m^2_{\chi_R}(t_3)|&\nonumber\\
=\ |m^2_{\chi_R}(t_2) +
m^2_{\chi_L}(t_3)|=\ |m^2_{ \chi_R}(t_3) +
m^2_{\chi_L}(t_2)|&\nonumber\\
\label{sdee}
\eeqa
The number of $h_1$, $h_2$ fields in the $bc^{\star}$ 
intersection is given by the intersection number of the 
$b$, $c^{\star}$ branes in the first
\footnote{Note that in this section
we imposed from the start that the  $h_1$, $h_2$
Higgs are present} tori for both PS-A, PS-B models,
\beq
n_{h^{\pm}}^{b c^{\star}}\
\stackrel{PS-A}{=}\ |\epsilon(m_c^1 - m_b^1)|,
\label{nadou1}
\eeq
\beq
n_{h^{\pm}}^{b c^{\star}}\
\stackrel{PS-B}{=}\ \beta_1|(n_b^1 + n_c^1)|.
\label{ser1}
\eeq
\newline
A comment is in order. Because the number of the electroweak
bidoublets in the PS-A models depends on the difference
$|m_b^1-m_c^1|$, it is more natural to solve the remaining tadpole
constraint (\ref{psatad1})
e.g. by making the choise
\beq
m_b^1 -m_c^1 = -( \b_2)(2 n_a^2 + n_d^2).
\label{remnain}
\eeq
Hence, e.g. by choosing $n_a^2=1,\;n_d^2=2,\;\b_2 =1/2$,
we get the constraint
\beq
|m_b^1-m_c^1| = \ 2,
\label{natural}
\eeq 
effectively choosing two electroweak Higgs bidoublets present.
Within this choise a consistent numerical set of wrappings
will be, we choose $\epsilon = {\tilde \epsilon}=1$, $m_b^1=-3$,
$m_c^1=-1$
\begin{center}
\beqa
N_a =4&(0,\ \ 1)(1, \ 3/2)(1,  1/2) \nonumber\\
N_b =2&(-1, \  -3)(2,\ \ 0) (1, \ 1/2) \nonumber\\
N_c =2&(1,\ \ -1)(2,\  \ 0)(1, -1/2) \nonumber\\    
N_d =1&(0,\ \ 1)(2, \ \ 3)(-2, \ 1)
\label{consist123}
\eeqa 
\end{center}

\subsection{Imposing $N=1$ SUSY on Intersections}

In this section, we will demand that certain sectors
respect $N=1$ supersymmetry. The reasons for doing so will become
absolutely clear
in the next section.
Up to this point the massless spectrum
of the PS-A, PS-B classes of models is that already described in
table (1).
In order for $N=1$ SUSY to be preserved at some intersection
between two branes  $\alpha$, $\beta$ we need to satisfy
$\pm \vartheta_{ab}^1 \pm \vartheta_{ab}^2 \pm \vartheta_{ab}^3$ for
some choise of signs, where
$\vartheta_{\alpha \beta}^i$, $i=1,2,3$
are the relative angles of the branes $\alpha$, $\beta$ across the
three 2-tori.

A Majorana mass term for neutrinos is absent
for PS-A, PS-B models when their massless spectrum is only the one given in
table (1).
This problem will disappear once we impose SUSY on intersections.
That will have as an effect the appearance of the massless scalar
superpartners
of the ${\bar F}_R$ fermions, the ${\bar F}_R^H$'s,
allowing a dimension 5
Majorana mass term for $\nu_R$, $F_R F_R {\bar F}_R^H {\bar F}_R^H$.

$\bullet$ {\em PS-A models}

We demand that the sectors $ac$, $dd^{\star}$ respect $N=1$ supersymmetry.
The conditions for $N=1$ SUSY on the sectors $ac$, $dd^{\star}$ are
respectively:
\beq
\pm (\frac{\pi}{2} +\ {\tilde \vartheta}_1) \ \pm  {\vartheta}_2 \
\pm  2\vartheta_3
\ = 0,
\label{condo1}
\eeq       
\beq
\pm \pi \ \pm \ 2 {\tilde \vartheta}_2 \ \pm 2 \vartheta_3 \ = 0
\label{condo2}
\eeq
These conditions can be solved by the choise, respectively,
\beq
ac \rightarrow (\frac{\pi}{2} + \ {\tilde \vartheta}_1) \ + \vartheta_2 \
- 2\vartheta_3 \ = 0,
\label{condo10}
\eeq       
\beq
dd^{\star} \rightarrow -\pi \ + \ 2 {\tilde \vartheta}_2 \ + \ 2 \vartheta_3 \ = 0
\label{condo20}
\eeq
and thus may be solved by the choise \footnote{We have set
$U^{(i)}= \frac{R_2^{(i)}}{R_1^{(i)}}$, $i=1, 2,3$}
\beq
-{\tilde \vartheta}_1 \ = {\tilde \vartheta}_2 =\ \vartheta_2 \ = \vartheta_3 \ = \frac{\pi}{4},
\label{solv1}
\eeq
effectively giving us
\beq
m_c^1 \ U^{(1)}= \ \frac{6 \b_2}{n_d^2} U^{(2)} = \
\frac{3 \b_2}{n_a^2} U^{(2)} = \  
  \frac{1}{2}U^{(3)} = \ \frac{\pi}{4}.
\label{condo3}
\eeq
The latter condition implies
\beq
2 n_a^2 =\ n_d^2.
\label{constra1}
\eeq
A set of wrapping numbers consistent with this constraint can be seen
in (\ref{consist123}).

By imposing $N=1$ SUSY on sectors $ac$, $dd^{\star}$ a massless scalar
partner appears in each sector. They are the massless scalar superpartner
of the fermions ${\bar F}_R$, $s_L$, namely the ${\bar F}_R^H$, $s_L^H$
respectively.
An additional feature of, see (\ref{condo3}), SUSY on intersections is that
the complex structure moduli $U^i$ takes specific values, 
decreasing the degeneracy of moduli
parameters in the theory.

A comment is in order.
If we list \footnote{see the 1st reference of \cite{cim}.} 
the vectors $ r$ describing a SUSY  
where we defined 
\beq
\begin{array}{c}
 r_0 = \pm \oh (+-+-) \nonumber \\
 r_1 = \pm \oh (++--) \nonumber\\
 r_2 = \pm \oh (-++-) \nonumber\\
 r_3 = \pm \oh (----) 
\end{array}
\label{r}
\eeq
then
the different SUSY's preserved by the branes 
$a$, $c$, $d$, $d^{\star}$ with the orientifold 
plane can be shown in table (\ref{higg}). As the intersection kl between branes
k and l will preserve the common supersymmetries that the branes k and l share with the orientifold plane it is manifest from table (\ref{r}) that the
sectors $ac$, $dd ^{\star}$ preserve $N=1$ SUSY.
Also notice that the $c$-brane preserves a $N=2$
SUSY.

\begin{table} [htb] \footnotesize
\renewcommand{\arraystretch}{1}
\begin{center}
\begin{tabular}{||c|c||c|c|c|}
\hline
\hline
Brane & $\theta_a^1$ & $\theta_a^2$ & $\theta_a^3$ & SUSY preserved\\
\hline\hline
$a$ & $\frac{\pi}{2}$  & $\frac{\pi}{4}$ & $\frac{\pi}{4}$ & $r_2$\\
\hline
$c$  & $-\frac{\pi}{4}$  & $0$ & $-\frac{\pi}{4}$ & $r_1$, $r_2$\\\hline\hline
$d$ & $\frac{\pi}{2}$  & $\frac{\pi}{4}$ & $\frac{3\pi}{4}$ & $r_1$\\
\hline
$d^{\star}$  & $\frac{3\pi}{2}$  & $\frac{7\pi}{4}$ & $\frac{5\pi}{4}$ 
& $r_1$\\\hline\hline
\end{tabular}
\end{center}
\caption{\small Angle content for branes participating supersymmetric 
sectors of PS-A models. The supersymmetry that is preserved by each brane with the O$_6$-plane is shown.
\label{higg}}
\end{table}

$\bullet$ {\em PS-B models}

In these models there is no $dd^{\star}$ sector, so we impose
$N=1$ SUSY on sector $ac$ only.
The condition for $N=1$ SUSY reads
\beq
\pm {\tilde \vartheta}_1 \ \pm \vartheta_2 \ \pm (\vartheta_3 +
{\tilde \vartheta}_3) = \ 0
\label{psbsus2}
\eeq
and is solved by
\beq
 {\tilde \vartheta}_1 \ + \vartheta_2 \ - (\vartheta_3 +
{\tilde \vartheta}_3) = \ 0
\label{psbsus2}
\eeq
with
\beq
\frac{U^{(1)}}{U^{(3)}} = \ \frac{3\rho^2 n_c^1}{2 \beta_1},\; \
\frac{U^{(2)}}{U^{(3)}} = \ \frac{3\rho^2 n_a^2}{2 \beta_2}.
\label{ok1}
\eeq

\subsection{Global symmetries}

Proton decay is one of the most important problems 
of grand unifies theories. In the standard versions of left-right 
symmetric PS models this problem could 
is avoided
as B-L is a gauged symmetry but the problem persists in 
baryon number violating operators of sixth order, contributing to proton 
decay. In our models, PS-A or PS-B, proton decay is absent as baryon number
 survives as a
global symmetry to low energies.
That provides for an explanation for the origin of proton stability
in general brane-world scenarios.

Clearly $Q_a = 3 B + L$ and the baryon B
is given by
\beqa
B = \frac{Q_a + Q_{B-L}}{4}.&  
\label{ba1}
\eeqa

As in the usual Pati-Salam model if the
neutral component of $H_1$ (resp. $H_2$), $\nu_H$,
assumes a vev, e.g $<$$\nu^H$$>$, then
the initial gauge symmetry,
$SU(4) \times SU(2)_L \times SU(2)_R
\times U(1)_a \times U(1)_b \times U(1)_c \times U(1)_d$,
  can break to the
standard model gauge group
$SU(3) \t U(2) \t U(1)_Y$ augmented by the non-anomalous $U(1)$ symmetry
$Q^l$. Lets us examine if it would be possible  
to break the extra $U(1)$ by appropriate Higgsing:

$\bullet$ {\em PS-A models}

In those models, by imposing SUSY on sector $dd^{\star}$ we
have the appearance of the scalar superpartner 
of $s_L$, the
${\tilde s }_L$ with the same multiplicity.
A linear combination of the $24\b^2$ singlets ${\tilde s }_L$
gets charged under the anomaly free
$U(1)$ symmetry (\ref{survi1}) and thus breaks the PS-A models to exactly
the much wanted SM gauge group structure,
$SU(3) \otimes SU(2) \otimes U(1)_Y$.
\newline
Note that it is
necessary on phenomenological grounds to break the extra non-anomalous
$U(1)$ (\ref{survi1}) that survives massless to low energy, as the surviving
gauge symmetry should be only of the observable standard model.
Its breaking
may be welcome as it provides the low energy standard 
model fermions with a flavour symmetry.
In the case of the non-anomalous $U(1)$ (\ref{survi1}) we deal with
these models
all SM fermions are not charged under it. 
Note that the extra non-anomalous $U(1)$ 
has some important phenomenological properties.
In particular it does not charge the PS symmetry breaking  Higgs 
 scalars $H_1$, $H_2$ thus avoiding the appearance of axions.
Note that the only issue remaining
is how we can give non-zero masses to all fermions of
table (1) beyond those of SM.

$\bullet$ {\em PS-B models}

In this case, even by imposing SUSY on intersections it is not possible
to create the Higgs particle with the right $U(1)$ charges that could
break the extra non-anomalous $U(1)$ symmetry to the SM itself.
 \newline
 A comment is in order.
 We note that the $F_R^H$ scalars coming from the $ac$ sector
 could be used as Higgs scalars that can break the PS left-right
 symmetry at the $M_{GUT}$ scale.
In this case it is not necessary to use the $H^{\pm}$ scalars as PS
breaking Higgses.

Also the analysis of the Higgs sector and neutrino
couplings (that follows) are
independent of the choises of extra $U(1)$'s, (\ref{survi1}),
(\ref{asda12}), (\ref{asda14}).

\section{Neutrino couplings and masses}

The analysis of neutrino masses that follows is valid for
both PS-A, PS-B models.
However, as we will see later in this subsection the class of PS-B models
have some shortcomings, e.g. the
fermions  $\chi_L$, $\chi_R$ could not get a mass.

On the contrary, the class of PS-A models has some remarkable features.
Namely, all extra fermions apart from SM one's get a mass and disappear
from the low energy spectrum. The only particles with light mass
close to the electroweak scale are those of fermions $\chi_L$. We note
that the fermions $\chi_L$, $\chi_R$ is a general prediction
of general left-right symmetric models in intersecting brane models
of type I strings and the mechasnism of making them
massive was unknown. Here, we find a way for giving them a mass in the
context of PS models.

In intersecting brane worlds trilinear Yukawa couplings between
the fermion states
$F_L^i$, ${\bar F}_R^j$ and the Higgs fields $H^k$ arise from the 
stretching of the worldsheet between the three D6-branes which cross 
at those intersections.
Its general form for a six dimensional torus is in the
leading order \cite{luis1},
\beq
Y^{ijk}=e^{- {\tilde A}_{ijk}},
\label{yuk1}
\eeq
where ${\tilde A}_{ijk}$ is the worldsheet area connecting the three vertices.
The areas of each of the two dimensional torus involved in
this interaction is typically of order one in string units.
To simplify matters we can without loss of generality asssume
that the areas of the second and third tori are close to zero.
In this case, the area of the full Yukawa coupling (\ref{yuk1})
reduces to
\beq
Y^{ijk}= e^{-\frac{R_1 R_2}{a^{\prime}} A_{ijk}},
\label{yuk12}
\eeq
where $R_1$, $R_2$ the radii and 
$ A_{ijk}$ the area of the two dimensional tori in the first complex
plane.  For a dimension five interaction term, like those involved in
the Majorana mass term for the right handed neutrinos the interaction term
is in the form
\beq
Y^{lmni}= e^{- {\tilde A}_{lmni}},
\label{yuk14}
\eeq
where ${\tilde A}_{lmni}$ the worldsheet area connecting the four interaction
vertices. Assuming that the areas of the second and third tori
are close to zero, the four term coupling can be approximated as
\beq
Y^{ijk}= e^{-\frac{R_1 R_2}{a^{\prime}} A_{lmni}},
\label{yuk15}
\eeq
where the area of the $A_{lmni}$ may be of order one in string units.
\newline
The full Yukawa interaction for the chiral spectrum of the PS-A, PS-B models
reads :
\beq
\lambda_1 F_L \ {\bar F}_R \ h 
+\ \lambda_2 \frac{F_R {F}_R {\bar F}_R^H {\bar F}_R^H }{M_s},  
\label{era1} 
\eeq
where
\beqa
\lambda_1 \equiv e^{-\frac{R_1 R_2 A_1}{\alpha^{\prime}}},&
\lambda_2 \equiv e^{-\frac{R_1 R_2 A_2}{\alpha^{\prime}}}.
\label{aswq123}
\eeqa
and the Majorana coupling involves the massless scalar \footnote{Of order 
of the string scale.}  
partners ${\bar F}_R^H$ 
of the antiparticles ${\bar F}_R$. This coupling is unconventional, in the
sense that the ${\bar F}_R^H$ is generated by imposing SUSY on an
sector of a non-SUSY model.
 We note the
 presence of $N=1$ SUSY at the sector $ac$. As can be seen by comparison
 with (\ref{na368}) 
 the ${\bar F}_R^H$ has a neutral direction that receives the vev $<H>$.   
There is no restriction for the vev of $F_R^H$ from first
principles and can be anywhere between the scale of elecrtoweak symmetry
breaking and $M_s$.
\newline
The Yukawa term 
\beqa
F_L {\bar F}_R h,&& h=\{h_1, h_2\},
\label{yukbre}
\eeqa
is responsible for the electroweak 
symmetry breaking. This term is responsible for 
giving Dirac masses to up quarks and
neutrinos.
In fact, we
get
\beq
\lambda_1 F_L {\bar F}_R h  \rightarrow (\lambda_1 \  \upsilon)
(u_i u_j^c + \nu_i N_j^c) + (\lambda_1 \  {\tilde \upsilon}) 
\cdot (d_i d_j^c + e_i e_j^c),  
\label{era2}
\eeq
where we have assumed that 
\beq
<h>= \left(
\begin{array}{cc}
\upsilon  & 0 \\
0 & {\bar \upsilon}
\label{era41}
\end{array}
\right)
\label{finalhiggs}
\eeq
We observe that the model gives non-zero tree level masses 
to the fields present. 
These mass relations may be retained at tree level only, since as the model
has a non-supersymmetric fermion spectrum, it
breaks supersymmetry on the brane, it will receive higher order 
corrections.  
It is interesting that from (\ref{finalhiggs}) we derive the GUT relation 
\cite{ellis}
\beq
m_d =\ m_e \ .
\label{gutscale}
\eeq
as well the ``unnatural''
 \beq
m_u =\ m_{N^c \nu} \ .
\label{gutscale1}
\eeq 

In the case of neutrino masses, 
the  ``unnatural'' (\ref{gutscale1}), associated
to the $\nu - N^c$
mixing,
is modified due to the presence of the Majorana term in (\ref{era1})
leading to a see-saw mixing type neutrino mass matrix in the form 
\beqa
\left(
\begin{array}{cc}
\nu&N^c 
\end{array}
\right)\times
 \left(
\begin{array}{cc}
0  & m \\
m & M
\label{era4}
\end{array}
 \right)
\times
\left(
\begin{array}{c}
\nu\\
N^c 
\label{era5}
\end{array}
\right),
\label{er1245}
\eeqa
where
\beq
m= \lambda_1  \upsilon.
\label{eigen1}
\eeq
After diagonalization
the neutrino mass matrix gives us two eigenvalues,
the ``heavy'' eigenvalue
\beq
m_{heavy} \approx M =\ \lambda_2 \frac{<H>^2 }{M_s},
\label{neu2}
\eeq
corresponding to the 
interacting right handed neutrino and 
the ``light'' eigenvalue
\beq
m_{light} \approx \frac{m^2}{M} =\ \frac{\lambda_1^2}{\lambda_2  }
\times\frac{\upsilon^2 \ M_s  } { <H>^2} 
\label{neu1}
\eeq
corresponding
to the interacting left handed neutrino. Note that 
the neutrino mass matrix is of
the type of an extended Frogatt- Nielsen mechanism \cite{fro} mixing
light with heavy states.

\begin{table} 
%[htb] \footnotesize
%\renewcommand{\arraystretch}{2.5}
\begin{center}
\begin{tabular}{|c||c|c|c|c|c|c|c|c|c|}
\hline
\hline
$M_s$ (GeV) & $10^3$  & $650$  & $600$  & $550$ & $500$ &
$450$ & $400$ & $350$  & $246$ \\\hline
$\upsilon^2/M_s$ (GeV) & $60$ & $93.1$ & $100.86$ & $110.03$ &
$121.03$ & $134.48$ & $151.29$ & $172.9$ & $246$ \\\hline
\end{tabular}
\end{center}
\caption{\small
Observe that the string scale cannot be at the TeV but lower.
Restricting the masses,
$\upsilon^2/M_S$, of left
handed fermion doublets
$\chi_L $ to values greater than 90 GeV and up to 246 GeV,
``pushes" the string scale to
values less than 650 GeV. The lower mass limit of $\chi_L$ pushes
the $M_s$ to maximum value.
\label{luk}}
\end{table}

\begin{table}
%[htb] \footnotesize
%\renewcommand{\arraystretch}{2.5}
\begin{center}
\begin{tabular}{|c|c|c|c||c|c||c||c|c||c|}
\hline
\hline
$M_s$ GeV& $\lambda_2$ & $A_2$ & $\langle H \rangle $  GeV &  $A_1$ 
& $R_1 R_2$ & $m_{\nu_R}(\leq E)GeV$ & $m_{\nu_L}$ eV\\
\hline\hline
$600$   & $\rightarrow 1$ & $\rightarrow 0$ & $600$    
&  $0.7$ & $8$ & $600$  
& $0.1$  \\\hline
$600$   & $\rightarrow 1$ & $\rightarrow 0$ & $600$    
&  $0.79$ & $8$ & $600$  
& $1$  \\\hline
$600$   & $\rightarrow 1$ & $\rightarrow 0$ & $600$    
&  $0.96$ & $6$ & $600$  
& $10$ \\\hline
$500$   & $\rightarrow 1$  & $\rightarrow 0$ & $500$    
&  $0.80$ & $8$ & $500$ 
& $1$ \\\hline
$500$   & $\rightarrow 1$  & $\rightarrow 0$ & $500$    
&  $0.97$ & $6$ & $500$ 
& $10$ \\\hline\hline
$550$   & $0.906$  & $77 A_1$ & $500$    
&  $\neq 0$ & $ \neq 0$ & $453$  
& $1$ \\\hline
$550$   & $0.906$  & $125 A_1$ & $500$ 
 &  $\neq 0$ & $\neq 0$ & $453$  
& $10$ \\\hline
$550$   & $0.906$  & $142 A_1$ & $500$ 
 &  $\neq 0$ & $\neq 0$ & $453$  
& $0.1$ 
\\\hline
\end{tabular}
\end{center}
\caption{\small Choises of the neutrino mass parameters for the 
$SU(4)_c \times SU(2)_L \times SU(2)_R $ type I model,
giving us hierarchical values of neutrino masses between 0.1-10 eV 
in consistency with oscillation experiments.
The Majorana mass term for the right handed neutrinos
involves a massive scalar 
superpartner with mass of order of  
the string scale.
The top
row shows the neutrino mass hierarchy when $\langle H \rangle = M_s$ while the
bottom part when $\langle H \rangle < M_s$. The analysis is valid for 
PS-A, PS-B classes of models. 
\label{spectrum101}}
\end{table}

Values of the parameters giving us values
for neutrino 
masses between 0.1-10 eV, consistent
with the observed neutrino mixing in neutrino
oscillation measurements, are shown in table (\ref{spectrum101}). The 
nature of the parameters involved in the Yukawa couplings (\ref{yuk1}),
generate naturally the hierarchy between the neutrino masses in the models.

In fact the hierarchy of neutrino masses can be investigated further
by examining several different scenaria associated with a light 
$\nu_L$ mass. 
As can be seen in table \ref{spectrum101}
there are two main options that are available to us:

\begin{itemize}

\item {\em $<H>$ $=$ $|M_s|$}

A long as the equality is preserved a consistent hierarchy of
neutrino masses is easily obtained.
It is important to note that the string scale cannot be at the TeV
but as we will show later it is constrained from the existence of the light
doublets $\chi_L$, to be less than 650 GeV. For simplicity, in table
(\ref{spectrum101}) we examine values of $M_s$ less than 600 GeV.
As long as $<H>$ $=$ $|M_s|$, the value of the $\lambda_2$
coupling should take the value one. In this case,
the area $A_2$ should tend to zero in order to have a non-zero value
for the product radii $R_1 \cdot R_2$, e.g $R_1 \cdot R_2 \neq 0$.

\item  {\em $<H>$ $<$ $| M_s |$}

In this case the structure of the theory is enough to
constrain the ratio of the areas $A_1$, $A_2$ involved in the
couplings of the see-saw mechanism. Lets us look for
example at the top row of the lower half of the table (\ref{spectrum101}).
By substituting the values of $M_s$, $<H>$, $m_{\nu_L}$, $m_{\nu_R}$
in (\ref{neu2}), (\ref{neu1}), we get the constraint equations
\beq
m_{\nu_R} \rightarrow R_1 R_2 A_2 =\ 0.05,\;m_{\nu_L} \rightarrow  R_1 R_2 A_1 =\ 3.67
\label{constr}
\eeq
effectively determining the value of the ratio $A_1/A_2 = 77$ independently
of the value of the product moduli $R_1 R_2$.
We note that because of the special nature of (\ref{neu2})
it is possible given the values for the string and the PS
breaking scale to determine the maximum values of $\nu_R$'s such that the
product radii $R_1 R_2$ is positive. A range of values for $\nu_R$ masses is shown
in table (\ref{string}).

\end{itemize}

\begin{table} 
%[htb] \footnotesize
%\renewcommand{\arraystretch}{2.5}
\begin{center}
\begin{tabular}{|c||c|c|c||}
\hline
\hline
$M_s$ (GeV) & $ \langle H \rangle $ (GeV) & $M_{\nu_R}\ <$  E  (GeV) \\\hline
$600.0$  & $500.0$ & $416.7$   \\
$550.0$  & $500.0$ & $454.5$   \\
$500.0$  & $400.0$ & $320.0$   \\
$600.0$  & $400.0$ & $266.7$    \\
$650.0$  & $470.1$ & $500.0$ \\
$650.0$  & $600.0$ & $553.8$
\\\hline\hline
\end{tabular}
\end{center}
\caption{\small
Bounds on $\nu_R$ for PS models, given the
scales $M_s$, $ \langle H \rangle $ .
\label{string}}
\end{table}

Notice that we have investigated the neutrino massses corresponding
to the first generation. This result could be extended to
covers all three generations.

Several comments are in order:

$\bullet$ {\em PS-A models}

Our main objective in this part is to show that
all additional particles, appearing in table (1), beyond those of SM get a 
heavy mass
and disappear from the low energy spectrum.
The slight exception will be the light mass of $\chi_L$
which is of order of the electroweak symmetry breaking scale.

Lets us discuss this issue in more detail.
The left handed fermions $\chi_L$ receive a mass
from the coupling 
\beq
(1, 2, 1)(1, 2, 1) e^{-A}
\frac{<h_2>< h_2>< {\bar F}_R^H ><H_1>< {\bar s}_L^H> }{M_s^4}
\stackrel{A \rightarrow 0}{\sim}
\frac{\upsilon^2}{M_s}(1, 2, 1)(1, 2, 1)
\label{kasa1}
\eeq
explicitly, in representation form, given by
\beqa
(1, 2, 1)_{(0, 1, 0, -1)} \ (1, 2, 1)_{(0, 1, 0, -1)} 
<(1, {\bar 2}, {\bar 2})_{(0, -1, -1, 0)}> \
<(1, {\bar 2}, {\bar 2})_{(0, -1, -1, 0)}> &\nonumber\\
\times \ <({\bar 4}, 1, 2)_{(-1, 0, 1, 0)}> \
<(4, 1, 2)_{(1, 0, 1, 0)}> \ 
<{\bf 1}_{(0, 0, 0, 2)}> 
\label{kasa111}
\eeqa
where we have included the leading contribution of the 
worksheet area connecting the
seven vertices. In the following for simplicity reasons we will set the 
leading contribution of the different couplings to
one (e.g. area tends to zero).
Altogether, $\chi_L$  
 receives a low mass of order $\upsilon^2/M_s$. 
Because there are no experimentally observed charged fermions, as can 
be seen for $e^{+}e^{-}$ interactions \footnote{I thank 
Luis Ib\'a\~nez for this comment.}, below 
90 GeV, by lowering the string scale below 1 TeV, in fact below 650 GeV,
we can push the $SU(2)_L$ fermions $\chi_L$ in the range between
 90 GeV and the scale of
 electroweak symmetry breaking. A range of values
showing different values of the string scale in connection
to $\chi_L$ masses
is shown in table (\ref{luk}).

Also, the $\chi_R$ doublet fermions receive heavy masses
in the following way.
The mass term
\beq
(1, 1, 2)(1, 1, 2)\frac{< H_2 > < F_R^H>< {\bar s}_L^H>}{M_s^2}
\label{real2}
\eeq
can be realized.
In explicit representation form
\beq
 (1, 1, 2)_{(0, 0, 1, 1)}  \ (1, 1, 2)_{(0, 0, 1, 1)} \
< ({\bar 4}, 1, {\bar 2})_{(-1, 0, -1, 0)}> \
< ({4}, 1, {\bar 2})_{(1, 0, -1, 0)}>  \
<{\bf 1}_{(0, 0, 0, 2)}>
\label{real200}
 \eeq
With vev's $<H_2> \sim <F_R^H>  \sim M_s$,
the mass of $\chi_R$ is of order $<s_L^H> /M_s$.
We note that in principle the vev of $s_L^H$, setting the 
scale of breaking of the extra anomaly free $U(1)$ could be anywhere 
between $<\upsilon>$ and $M_s$. However, since $M_s$ is constrained to be less
or equal to 650 GeV, given the proximity of the 
intermediate scale $s_L^H$ and the string scale, we could
suppose for the rest of this
work that $s_L^H \sim M_s$.
However, in principle the vev of $s_L^H$ can be anywhere between
$246$ GeV and $M_s$, the latter up to $650$ GeV.

The 10-plet fermions $z_R$ receive
 a heavy mass of order $M_s$ from the coupling
\beq
(10, 1, 1)(10, 1, 1)\frac{<{\bar F}_R^H ><{\bar F}_R^H>< H_2>< H_2>}{M_s^3},
\label{10plet}
\eeq
where we have used the
tensor product representations for $SU(4)$,
$10 \otimes 10 = 20 + 35 + 45$,
$20 \otimes {\bar 4} = {\bar 15 } + {\bar 20}$, 
${\bar 20} \otimes {\bar 4} = {\bar 6 } + 10$,
$10 \otimes {\bar 4} =  4  + 36$, $4 \otimes {\bar 4} = 1 + 15$.
Explicitly, in representation form, 

\beqa
(10, 1, 1)_{(2, 0, 0, 0)} (10, 1, 1)_{(2, 0, 0, 0)}
<({\bar 4}, 1, 2)_{(-1, 0, 1, 0)}> \
<({\bar 4}, 1, {2})_{(-1, 0, 1, 0)}>
&\nonumber\\
\times  \
<({\bar 4}, 1, {\bar 2})_{(-1, 0, -1, 0)}> \
<({\bar 4}, 1, {\bar 2})_{(-1, 0, -1, 0)}>
\label{10pletagain}
\eeqa

The 6-plet fermions, $\omega_L$, receive a mass term of
order $M_s$  from the
coupling, e.g. for $\omega_L$
\beq
({\bar 6}, 1, 1)({\bar 6}, 1, 1)
\frac{<H_1 ><{F}_R^H>< H_1>< {F}_R^H>}{M_s^3}
\label{6plet}
\eeq
where we have made use of the $SU(4)$ tensor products
$6 \otimes 6 = 1 + 15 + 20$, $ 4 \otimes 4 = 6 + 10$.
Explicitly, in representation form,
\beqa
({\bar 6}, 1, 1)_{(-2, 0, 0, 0)}     \ ({\bar 6}, 1, 1)_{(-2, 0, 0, 0)} 
<(4, 1, 2)_{(1, 0, 1, 0)} > \ <(4, 1, 2)_{(1, 0, 1, 0)} > &\nonumber\\
\times \ <(4, 1, {\bar 2})_{(1, 0, -1, 0})> \ <(4, 1, {\bar 2})_{(1, 0, -1, 0})>
\label{6plet}
\eeqa

Finally, the singlet fermions $s_L$ receive a mass of order $M_s$ from the
coupling
\beq
{\bar s}_L{\bar s}_L \frac{<s_L^H> <s_L^H>}{M_s}
\label{1plet}
\eeq
Thus only the chiral fermion content of the SM fermions remains at $M_Z$.

$\bullet$ {\em PS-B models}

While the neutrino sector of those models can give small
masses to neutrinos, the main shortcoming of the models is that
the fermion doublets $\chi_L$, $\chi_R$ remain massless
down to the electroweak scale in contrast with the observed low
energy phenomenology. Also 
the $U(1)$ symmetry (\ref{hyper}) survives unbroken to low energies.
 Thus PS-B models are phenomenologically
not interesting in this respect.

\section{Conclusions}

In this work, we have presented the first
examples of four dimensional string grand unified models 
that can give at low energy exactly the observable standard
model spectrum and gauge interactions. These models,
characterized as PS-A class in this work,
are based on the Pati-Salam gauge group
$SU(4)_C \times SU(2)_L \times SU(2)_R$ and 
are derived from D6-branes intersecting at non-trivial angles in four
dimensional
type I
compactifications on a six dimensional orientifolded torus.
The models have their quarks and leptons accommodated in
three generations,
and possess some remarkable features.
Among them we mention that the models give some answers as
matter as it
concerns one of the most difficult aspects of gauge hierarchy, 
apart from the hierarchy of scales, that is the
smallness of neutrino masses.

In this case it is particularly easy for the theory to accommodate
a neutrino mass hierarchy between 0.1-10 eV consistent with oscillation
measurements.

Through out the paper we distinguished the different PS GUT solutions
according
to if the tadpoles admit or not exotic, antisymmetric and symmetric,
reprsesentations of $U(N_a)$ groups coming from brane-orientifold
image brane,
$\alpha \alpha^{\star}$, sectors.
In this way, PS-A models, that give exactly the SM at low energies,
possess $\alpha \alpha^{\star}$ sectors. On the contrary, PS-B models
which don't admit
$\alpha \alpha^{\star}$ sectors, failed to produce just the SM at
low energies. However, some important conclusions were derived from the
study of PS-B models.
We got an interpretation of the appearance of multi-brane wrapping
in intersecting branes. It appears that, since in the absence of a 
stringy Higgs effect no more additional $U(1)$'s may be introduced,
the additional $U(1)$'s can be absorbed into a trivial field
redefinition of the non-anomalous $U(1)$, 
surviving the Green-Schwarz mechanism  
at low energies.
Moreover,
colour triplet Higgs couplings that could couple to quarks and leptons 
and cause a problem to proton decay are absent in all classes of models.
Proton is stable as baryon number survives as 
global symmetry to low energies.

We should note that
a hint of motivation from searching for Grand Unified models (GUTS), 
comes from the fact that very recently, there is evidense
from neutrinoness beta-decay, even though not conclusive,
 for the existense of  non-zero Majorana masses for neutrinos
and lepton number violation \cite{decay}.
\newline
Despite the fact,that the models we examined are 
free of RR tadpoles and, if the angle
stabilization conditions of Appendices I, II hold, free of tachyons, 
they will always have NSNS tadpoles 
that cannot all be removed.
The closed string NSNS tadpoles can be removed by freezing
the complex moduli
to discrete values \cite{blume}, or by redefining the background 
in terms of 
wrapped metrics \cite{nsns}.
However, a dilaton tadpole will always remain that could in principle 
reintroduce tadpoles in the next leading order.
A different mechanism, involving different type I compactification
backgrounds to the one
used in this article, that could avoid global tadpoles
was described in \cite{abla}.
We note that for PS-A models the complex structure moduli \footnote{
The
K\"ahler moduli could be
fixed from its value at the string scale, using
for example the product radii in (\ref{constr}) but that would mean too
large fine tuning for our theory to be naturally existent.}
can be fixed
to discrete values, e.g. see (\ref{condo3}).

One point that there was no obvious stringy solution with general 
orientifolded
six-torus compactifications   
is that these models do not offer an apparent explanation
for keeping the string scale low \cite{antoba}, e.g to 1-100 TeV region.
This aspect of the hierarchy
that makes the Planck scale large, while keeping the string scale low,
 by varying the radii of the 
transverse directions \cite{antoba} does not apply here, as there are no
transverse torus directions simultaneously to all D6-branes \cite{tessera}. 
A possible solution, even though such manifolds are not known,
was suggested in \cite{luis1}, could involve
cutting a
ball, to a region away from the D6-branes, and gluing a throat connecting
the T6 torus to a large volume manifold. 
However, in this work we suggested an alternative mechanism
that keeps the string scale $M_s$ low. In particular the existence
of the light weak doublets in the PS-A models with a mass of order up to
246 GeV, makes a definite prediction for a low string scale 
in the energy range less than 650 GeV.   
That effectively, makes the PS-A class of D6-brane models directly 
testable to present or feature accelerators.

The general structure of the GUT models with PS structure 
presented in this article
contains at low energy the standard model augmented by a non-anomalous 
$U(1)$ symmetry. For the PS-A class this additional $U(1)$ 
was broken by extra singlets that were created after modifying certain
non-SUSY sectors such that they preserve $N=1$ supersymmetry.
Thus it appears that the model has $N=1$ SUSY sectors even though
overall is a non-SUSY model. Furthermore, the broken, anomaly free
$U(1)$ symmetry, charges the fermions of the standard model with an
interesting flavour symmetry.

String models, similar
to present, without the presence of exotic matter and/or additional gauge 
group content (from gravity mediating ``hidden'' sectors) a low energies,
has
appeared
in \cite{louis2, kokos}, where however,
the authors were
able to have just the standard model at low energies without using
a grand unified structure.

Also, it will be interesting to extend the methods employed in this article,
to other GUT groups.
Summarizing, in the present work, we have shown that 
 we can start from a  
realistic Pati-Salam structure at the string scale and derive
the first GUT string examples with exactly the observable standard
model at low energies.

\begin{center}
{\bf Acknowledgments}
\end{center}
I am grateful to Ignatios Antoniadis, Daniel Cremades, 
Luis Ib\'a\~nez, Fernado Marchesano
and Angel Uranga,
for useful discussions and comments. 
In addition, I would like to thank the CERN Theory Division for its warm 
hospitality during
the completion of this work.

\newpage

\section{Appendix I}

In the appendix we list the conditions, mentioned in subsection (4.1), 
  under which the PS-A model D6-brane configurations
of tadpole solutions of table (\ref{spectruma101}), are tachyon free. 
Note that 
the conditions are expressed in terms of the angles
defined in (\ref{angPSA}).

\beqa
\begin{array}{ccccccc}
-\vartheta_1 &+& \vartheta_2 &+& 2 {\vartheta}_3 &\geq& 0\\
(-\frac{\pi}{2}+{\tilde \vartheta}_1) &+& \vartheta_2 &+& 
2{\vartheta}_3 &\geq &0 \\
(\frac{\pi}{2}-\vartheta_1) &+& \pi-{\vartheta}_2 &+& \pi-2 {\vartheta}_3 &\geq& 0\\
(-\frac{\pi}{2}-{\tilde \vartheta}_1) &+&\pi-{\vartheta}_2 &+&\pi- 
2{\vartheta}_3 &\geq& 0 
\\\\
\vartheta_1 &-& \vartheta_2 &+&2 {\vartheta}_3&\geq& 0  \\
(\frac{\pi}{2}-{\tilde \vartheta}_1) &-& \vartheta_2 & +&2{\vartheta}_3 &\geq& 0\\
(-\frac{\pi}{2}+{\vartheta}_1) &+&(-\pi+{\vartheta}_2) & 
+&(\pi-2{\vartheta}_3) &\geq& 0\\
(\frac{\pi}{2}+{\tilde \vartheta}_1) &+&(-\pi+{\vartheta}_2) 
& +&(\pi-2{\vartheta}_3) &\geq& 0\\\\
{\vartheta}_1 &+& \vartheta_2 & -&2{ \vartheta}_3 &\geq& 0\\
(\frac{\pi}{2}-{\tilde \vartheta}_1) &+& \vartheta_2 & -&2{ \vartheta}_3 &\geq& 0\\
(-\frac{\pi}{2}+{ \vartheta}_1) &+&(\pi-{\vartheta}_2) & +&
(-\pi+2{ \vartheta}_3) &\geq& 0\\
(\frac{\pi}{2}+{\vartheta}_1) &+&(\pi-{\vartheta}_2) & +&
(-\pi+2{\vartheta}_3) &\geq& 0\\
\label{free}
\end{array}
\eeqa

\newpage

\section{Appendix II}

In the appendix we list the conditions, mentioned in subsection (4.1), 
  under which the PS-B model D6-brane configurations
of tadpole solutions of table (\ref{spectrum8}), are tachyon free. Note that 
the conditions are expressed in terms of the angles
defined in (\ref{angulos}) and furthermore we have take into account that
$\vartheta_3 
= {\tilde \vartheta}_3$.

\beqa
\begin{array}{ccccccc}
-\vartheta_1 &+& \vartheta_2 &+& 2 {\tilde \vartheta}_3 &\geq& 0\\
-{\tilde \vartheta}_1 &+& \vartheta_2 &+& 2{\tilde \vartheta}_3 &\geq &0 \\
-\vartheta_1 &+& {\tilde \vartheta}_2 &+& 2 {\tilde \vartheta}_3 &\geq& 0\\
-{\tilde \vartheta}_1 &+&{\tilde \vartheta}_2 &+& 2{\tilde \vartheta}_3 &\geq& 0 
\\\\
\vartheta_1 &-& \vartheta_2 &+&2 {\tilde \vartheta}_3&\geq& 0  \\
{\tilde \vartheta}_1 &-& \vartheta_2 & +&2{\tilde \vartheta}_3 &\geq& 0\\
{\vartheta}_1 &-&{\tilde  \vartheta}_2 & +&2{\tilde \vartheta}_3 &\geq& 0\\
{\tilde \vartheta}_1 &-&{\tilde \vartheta}_2 & +&2{\tilde \vartheta}_3 &\geq& 0\\\\
{\vartheta}_1 &+& \vartheta_2 & -&2{\tilde \vartheta}_3 &\geq& 0\\
{\tilde \vartheta}_1 &+& \vartheta_2 & -&2{\tilde \vartheta}_3 &\geq& 0\\
{\tilde \vartheta}_1 &+&{\tilde  \vartheta}_2 & -&2{\tilde \vartheta}_3 &\geq& 0\\
{\tilde \vartheta}_1 &+&{\tilde  \vartheta}_2 & -&2{\tilde \vartheta}_3 &\geq& 0\\
\label{free1}
\end{array}
\eeqa

\newpage

\section{Appendix III}

The PS-B models appearing in table one can be proved that can be equivalent
to the ones created after assigning 
the alternative accommodation 
of fermions charges, of table \ref{alternative}, below :

\begin{table} 
[htb] \footnotesize
\renewcommand{\arraystretch}{1}
\begin{center}
\begin{tabular}{|c||c|c||c||c|c|}
\hline
 Intersection  & $SU(4)_C \times SU(2)_L \times SU(2)_R$ &
$Q_a$ & $Q_b$ & $Q_c$ & $Q_d$ \\
\hline
  $I_{ab^{\ast}}=3$ &
$3 \times (4,  2, 1)$ & $1$ & $1$ & $0$ &$0$ \\
   $I_{a c}=-3 $ & $3 \times ({\ov 4}, 1, 2)$ &
$-1$ & $0$ & $1$ & $0$\\
 $I_{bd^{\star}} = -12$ &  $12 \times (1, {\ov 2}, 1)$ &
$0$ & $-1$ & $0$ & $-1$ \\    
  $I_{cd} = -12$ &  $12 \times (1, 1, {\ov 2})$ &
$0$ & $0$ & $-1$ &$1$ \\    
\hline
\end{tabular}
\end{center}
\caption{\small Alternative accommodation of chiral spectrum for
 the $SU(4)_C \times 
SU(2)_L \times SU(2)_R$ type I PS-B models, discussed in the main boby of the paper,
 together with $U(1)$ charges.
\label{alternative}}
\end{table}
For the accommodation of Pati-Salam models with alternative 
fermion charges listed in table \ref{alternative},
the full solutions to the tadpole constraints are given by the following
tables :

\begin{table}[htb]\footnotesize
\renewcommand{\arraystretch}{2}
\begin{center}
\begin{tabular}{||c||c|c|c||}
\hline
\hline
$N_i$ & $(n_i^1, m_i^1)$ & $(n_i^2, m_i^2)$ & $(n_i^3, m_i^3)$\\
\hline\hline
 $N_a=4$ & $(1/\beta_1, 0)$  &
$(n_a^2, - \epsilon \b_2)$ & $(1/\rho, \frac{3\rho}{{2}})$  \\
\hline
$N_b=2$  & $(n_b^1, \epsilon \b_1)$ & $(1/\beta_2, 0)$ &
$(1/\rho, \frac{3\rho}{{2}})$ \\
\hline
$N_c=2$ & $(n_c^1, \epsilon \b_1)$ &   $(1/\beta_2, 0)$  & 
$(1/\rho, -\frac{3\rho}{{2}})$ \\    
\hline
$N_d=1$ & $(\alpha/{\beta_1}, 0)$ &  $(n_d^2,  \gamma\epsilon \b_2)$  
  & $(1/{\rho}, \frac{3\rho}{{2}})$  \\   
\hline
\end{tabular}
\end{center}
\caption{\small First class of solutions 
for alternative accommodation of fermion charges, of 
D6-branes wrapping numbers giving rise to the 
fermionic spectrum of the 
$SU(4)_C \times SU(2)_L \times SU(2)_R$ type I PS-B models of table (1).
The parameter $\rho$ takes the values $1, 1/3$, while there 
is an additional dependence on four integer parameters, 
$n_a^2$, $n_d^2$, $n_b^1$, $n_c^1$, the NS-background $\beta_i$ and 
the phase parameter $\epsilon = \pm 1$. Note the condition $\alpha \gamma=4$
and the positive wrapping number entry on the 3rd tori of 
the a-brane.  
\label{alte1}}
\end{table}

\newpage

\begin{table}[htb]\footnotesize
\renewcommand{\arraystretch}{1.5}
\begin{center}
\begin{tabular}{||c||c|c|c||}
\hline
\hline
$N_i$ & $(n_i^1, m_i^1)$ & $(n_i^2, m_i^2)$ & $(n_i^3, m_i^3)$\\
\hline\hline
 $N_a=4$ & $(1/\beta_1, 0)$  &
$(n_a^2, - \epsilon \b_2)$ & $(1/\rho, -\frac{3\rho}{{2}})$  \\
\hline
$N_b=2$  & $(n_b^1, \epsilon \b_1)$ & $(1/\beta_2, 0)$ &
$(-1/\rho, \frac{3\rho}{{2}})$ \\
\hline
$N_c=2$ & $(n_c^1, \epsilon \b_1)$ &   $(1/\beta_2, 0)$  & 
$(-1/\rho, -\frac{3\rho}{{2}})$ \\    
\hline
$N_d=1$ & $(\alpha/{\beta_1}, 0)$ &  $(n_d^2,  \gamma\epsilon \b_2)$  
  & $(1/{\rho}, -\frac{3\rho}{{2}})$  \\   
\hline
\end{tabular}
\end{center}
\caption{\small Second class of solutions, 
for alternative accommodation of fermion charges,
 of 
D6-branes wrapping numbers giving rise to the 
fermionic spectrum of the 
$SU(4)_C \times SU(2)_L \times SU(2)_R$ type I PS-B models of table (1).
The parameter $\rho$ takes the values $1, 1/3$, while there 
is an additional dependence on four integer parameters, 
$n_a^2$, $n_d^2$, $n_b^1$, $n_c^1$, the NS-background $\beta_i$ and the 
phase parameter $\epsilon = \pm 1$. Note the condition $\alpha \gamma=4$
and the positive wrapping number entry on the 3rd tori of 
the a-brane.  
\label{alte2}}
\end{table}

The surviving $U(1)$ anomalous in this case reads :
\beq
{\tilde Q}_l =\ (Q_b -\ Q_c) -(Q_a +\ Q_d),  
\label{alter1}
\eeq
where an identical set of wrapping number solutions to 
(\ref{numero2}) has been chosen.
The low energy theory is the standard model augmented by the global gauged
$U(1)$ ${\tilde Q}_l$.


\newpage

\begin{thebibliography}{99}
\bibitem{ena} For reviews, see, {\it e.g.},
F.~Quevedo,
``Lectures on superstring phenomenology,''
{\tt hep-th/9603074};
A.~E.~Faraggi,
``Superstring phenomenology: Present and future perspective,''
{\tt hep-ph/9707311};
Z.~Kakushadze, G.~Shiu, S.~H.~Tye and Y.~Vtorov-Karevsky,
``A review of three-family grand unified string models,''
Int.\ J.\ Mod.\ Phys.\ A {\bf 13}, 2551 (1998).

\bibitem{dio1}
A.~Sagnotti, in Cargese 87, {\it Strings  on Orbifolds}, ed. G.Mack et al.
(Pergamon Press, 1988) p. 521;
P.~Horava, \NPB327 (1989) 461; \PLB231 (1989) 251;
J.~Dai, R.~Leigh and J.~Polchinski, Mod. Phys. Lett. A4 (1989) 2073; 
R.~Leigh, Mod. Phys. Lett. A4 (1989) 2767;
 G.~Pradisi and A.~Sagnotti, \PLB216 (1989) 59 ;
M.~Bianchi and A.~Sagnotti, \PLB247 (1990) 517 ;
E.~Gimon and J.~Polchinski, Phys.Rev. D54 (1996) 1667, hep-th/9601038.
E.~Gimon and C.~Johnson, \NPB477  (1996) 715 , hep-th/9604129; 
A.~Dabholkar and J.~Park, \NPB477 (1996) 701, hep-th/9604178.

\bibitem{ibaba}
M.~Berkooz and R.~G.~Leigh, \NPB483 (1997) 187, hep-th/9605049;
Z.~Kakushadze, \NPB512 (1998) 221, hep-th/9704059;
Z.~Kakushadze and G.~Shiu, \PRD56 (1997) 3686, hep-th/9705163;
Z.~Kakushadze and G.~Shiu, \NPB 520 (1998) 75, hep-th/9706051;
L.E.~Ib\'a\~nez, JHEP 9807 (1998) 002, hep-th/9802103;
G.~Zwart, hep-th/9708040;
J. Lykken, E. Poppitz and S. Trivedi, 
``Branes with GUTs and supersymmetry breaking,''
\NPB 543  (1999) 105,
hep-th/9806080 ;
G. Aldazabal, A. Font, L.E. Ib\'{a}\~{n}ez and G. Violero,
 \NPB536 (1998) 29, ``D=4,N=1, Type IIB Orientifolds''
hep-th/9804026;
 G. Aldazabal,  L.~E.~Ib\'a\~nez, F. Quevedo and
A.~M.~Uranga, 
``D-branes at singularities: A bottom-up approach to the string
embedding  of the standard model,''
JHEP 0008:002, 2000, hep-th/0005067;
Z. Kakushadze and S. H. Tye,
``Three Generations in Type I Compactifications''
Phys. Rev. D58 (1998) 126001


\bibitem{tessera}R.~Blumenhagen, L.~G\"orlich, B.~K\"ors and D.~L\"ust,
``Noncommutative compactifications of type I strings on tori with magnetic
background flux'',
JHEP {\bf 0010} (2000) 006, {\tt hep-th/0007024};
``Magnetic Flux in Toroidal Type I Compactification'', Fortsch. Phys. 49
(2001) 591, hep-th/0010198 


\bibitem{dikomou}
L. Dixon, V. Kaplunovsky, and J. Louis, ``Moduli Dependence
of String Loop Corrections to Gauge Coupling Constants'', Nucl. 
Phys. B355 (1991) 649;\\
 C.~Kokorelis, ``String Loop Threshold Corrections for $N=1$
 Generalized Coxeter Orbifolds'',
 Nucl. Phys. B579 (2000) 267, hep-th/0001217;\\
D. Bailin, A. Love, W. Sabra, S. Thomas,  ``String Loop Threshold Corrections
for $Z_N$ Coxeter Orbifolds'', Mod. Phys. Let. A9 (1994) 67, hep-th/9310008;\\
 C.~Kokorelis, ``Gauge and
 Gravitational Couplings from Modular Orbits in
 Orbifold Compatifications'', Phys. Lett. B477 (2000) 313 , hep-th/0001062



\bibitem{antoba}
N.Arkadi-Hamed, S. Dimopoulos and G. Dvali, \PLB429 
(1998) 263; I. Antoniadis, N.Arkadi-Hamed, S. Dimopoulos and G. Dvali,
\PLB436 (1998) 257; I. Antoniadis and C. Bachas, \PLB450 (1999) 83



\bibitem{pente}
E.S. Fradkin and A.A. Tseytlin, \PLB 158 (1985) 316; A. Abouelsaood, 
C.G. Gallan, C.R. Nappi and S.A. Yost, \NPB 289 (1987) 599
C.~Bachas,
``A Way to break supersymmetry'',
{\tt hep-th/9503030}.

\bibitem{tessera1}
M.M. Sheikh-Jabbari,
``Classification of Different Branes at Angles'',
Phys.Lett. B420 (1998) 279-284, hep-th/9710121. 
H. Arfaei, M.M. Sheikh-Jabbari,
``Different D-brane Interactions'',
Phys.Lett. B394 (1997) 288-296, hep-th/9608167 
R. Blumenhagen, L. Goerlich, B. Kors,
``Supersymmetric Orientifolds in 6D with D-Branes at Angles'',
Nucl.Phys. B569 (2000) 209-228,
hep-th/9908130; 
R. Blumenhagen, L. Goerlish, B. K\"ors,
``Supersymmetric 4D Orientifolds of Type IIA with D6-branes at
Angles'',
JHEP 0001 (2000) 040,hep-th/9912204; 
S. Forste, G. Honecker, R. Schreyer,
``Supersymmetric $Z_N \times Z_M$ Orientifolds in 4D with
D-Branes at Angles'',
Nucl.Phys. B593 (2001) 127-154, hep-th/0008250;
Ion V. Vancea,
``Note on Four Dp-Branes at Angles'',
JHEP 0104:020,2001,
hep-th/0011251; H. Kataoka, M.Shimojo, ``$SU(3) \times SU(2) \times
U(1)$
Chiral models from Intersecting D4-/D5-branes'', hep-th/0112247;
G. Honecher, ``Intersecting brane world models from D8-branes on
$(T^2 \times T^4/Z_3)/\Omega {\cal R}_1$ type IIA orientifolds'',
hep-th/0201037

\bibitem{bele}M. Berkooz, M. R. Douglas, R.G. Leigh, 
``Branes Intersecting at Angles'',
\NPB480 (1996) 265, 
hep-th/9606139

\bibitem{eksi1}
M.~Bianchi, G.~Pradisi and A.~Sagnotti,
``Toroidal compactification and symme
try breaking in open string theories,''
Nucl.\ Phys.\ B {\bf 376}, 365 (1992)
 
\bibitem{eksi2}
Z.~Kakushadze, G.~Shiu and S.-H.~H.~Tye,
``Type IIB orientifolds with NS-NS antisymmetric tensor backgrounds,''
Phys.\ Rev.\ D {\bf 58}, 086001 (1998).
hep-th/9803141

\bibitem{eksi3}C. Angelantonj, \NPB 566 (2000) 126, 
``Comments on Open-String Orbifolds with a Non-Vanishing 
$B_{ab}$'', hep-th/9908064

\bibitem{tessera2}R. Blumenhagen, B. K\"ors and D. L\"ust,
``Type I Strings
with F and B-flux'',
JHEP 0102 (2001) 030, hep-th/0012156.

\bibitem{carlo}
R. Blumenhagen, L. G\"orlish, and B. K\"ors, ``Asymmetric Orbifolds, 
non-commutative geometry and type I string vaua'', 
Nucl. Phys. B582 (2000) 44, hep-th/0003024\\ 
C. Angelantoj and A. Sagnotti, 
``Type I vacua and brane transmutation'', hep-th/00010279;\\
C. Angelantoj, I. Antoniadis, E. Dudas and A. Sagnotti,
``Type I strings on magnetized orbifolds and brane transmutation'',
Phys. Lett. B489 (2000) 223, hep-th/0007090

\bibitem{luis1}
G.~Aldazabal, S.~Franco, L.~E.~Ib\'a\~nez, R.~Rabad\'an and
A.~M.~Uranga,
``D=4 chiral string compactifications from intersecting branes'',
J. Math. Phys. 42 (2001) 3103-3126,
{\tt hep-th/0011073};
G.~Aldazabal, S.~Franco, L.~E.~Ib\'a\~nez, R.~Rabad\'an and
A.~M.~Uranga,
``Intersecting brane worlds'',
JHEP {\bf 0102} (2001) 047, {\tt hep-ph/0011132}.

\bibitem{louis2}
L.~E.~Ib\'a\~nez, F.~Marchesano and R.~Rabad\'an,
``Getting just the standard model at intersecting branes''
JHEP, 0111 (2001) 002, {\tt hep-th/0105155};
L.~E.~Ib\'a\~nez,
``Standard Model Engineering with Intersecting Branes'',
hep-ph/0109082

\bibitem{kokos}C. ~Kokorelis, ``New Standard Model Vacua from Intersecting 
Branes'', hep-th/0205147


\bibitem{alda} L. F. Alday and G. Aldazabal, ``In quest of 
"just" the Standard Model on D-branes at a singularity'', 
hep-th/0203129

\bibitem{blume}R.~Blumenhagen, B.~K\"ors, D.~L\"ust and T. Ott,
``The Standard Model from Stable Intersecting Brane World Orbifolds'',
\NPB616 (2001) 3, hep-th/0107138

\bibitem{uran}M. Cvetic, G. Shiu, A. M. Uranga,
``Chiral four dimensional N=1 supersymmetric type 2A orientifolds from
intersecting D6 branes'', 
Nucl. Phys. B615 (2001) 3;\\
hep-th/0107166
``Three family
supersymmetric standard models from intersecting brane worlds'',
Phys. Rev. Lett. 87 (2001) 201801, hep-th/0107143;
M. Cvetic, P. Langacker, G. Shiu, 
``Phenomenology of A Three-Family Standard-like String Model'',
hep-ph/0205252


\bibitem{Pati-Salam}J. Pati and A. Salam, ``Lepton number as a 
fourth colour'',
Phys. Rev. D10 (1974) 275


\bibitem{agui} A. Aguilar at al. [LSND Collaboration], 
``Evidence for Neutrino Oscillations from the Observation of Electron
Anti-neutrinos in a Muon Anti-Neutrino Beam'', Phys. Rev. D64, 112007 
(2001), hep-ex/0104049; C. Athanasopoulos et al., 
``Evidence for $\nu_{\mu} -> \nu_e$ Oscillations from Pion Decay in Flight
Neutrinos'', Phys. Rev. C58, 2849 (1998),
nucl-ex/9706006

\bibitem{fro} C. D. Frogatt and H. B. Nielsen, 
``Hierarchy of quark masses, Cabibbo angles and CP violation'',
\NPB 147 (1979) 277





\bibitem{raya}M. GellMann, P. Ramond and R. Slansky, in {\em Supergravity}, ed.
F. van Nieuwenhuizen and D. Friedman, (North Holland, Amsterdam, 1979)p. 315;
T. Yanagida, {\em Proc. of the Workshop on on Unified Theory and the Baryon 
Number of the Universe}, KEK, Japan, 1979;
S. Weinberg, Phys. Rev. Lett. 43, 1566 (1979). 




\bibitem{kaku}
Z. Kakushadze, ``A Three-Family $SU(4)_c \times SU(2)_w \times U(1)$ Type I
Vacuum'',
Phys. Rev D58 (1998) 101901, hep-th/9806044

\bibitem{luis3} G. Aldazabal, L.~E.~Ib\'a\~nez, F. Quevedo,  
``Standard-like 
Models with Broken Supersymmetry from Type I String Vacua'' ,
{\tt hep-th/9909172}.

\bibitem{antokt}I. Antoniadis, E. Kiritsis 
and T. Tomaras, 
``D-brane Standard Model''
\PLB486 (2000) 186, hep-th/0111269

\bibitem{lr}G. K. Leontaris and J. Rizos,
``A Pati-Salam model from branes'',
 \PLB510 (2001) 295, hep-ph/012255

\bibitem{bere} D. Berenstein, V. Jejjala and R.G. Leigh,
 ``The Standard Model on a D-brane'',
hep-ph/0105042




\bibitem{anto}I. Antoniadis and G.K. Leontaris,
``A supersymmetric $SU(4) \times O(4)$ model'' \PLB216 (1989) 333;
I. Antoniadis, G. K. Leontaris and J. Rizos, ``
A three generation $SU(4) \times O(4)$ string model''
\PLB245 (1990)161;
G.K. Leontaris, \PLB372 (1996) 212, ``A String Model with 
$SU(4)\times O(4)\times [Sp(4)]_{Hidden}$ Gauge Symmetry'', 
hep-ph/9601337 


\bibitem{sa}A. Sagnotti,
``A Note on the Green - Schwarz Mechanism in Open - String Theories'',
 Phys. Lett. B294 (1992) 196, hep-th/9210127

\bibitem{giapo}A. Murayama and A. Toon,
``An $SU(4) \otimes SU(2)_L \otimes SU(2)_R$ string model
wit direct unification at the string scale'',
\PLB318 (1993)298,



\bibitem{ellis}M. S. Chanowitz, J. Ellis and M. K. Gailard,
``The price of natural flavour conservation in neutral weak interactions'',
\NPB 128 (1977) 506




\bibitem{iru}L.~E.~Ib\'a\~nez, R.~Rabad\'an and A. M. Uranga,
``Anomalous U(1)'s in Type I and Type IIB D=4, N=1 string vacua'',
Nucl.Phys. B542 (1999) 112-138;\\
M. Klein, ``Anomaly cancellation in D=4 N=1 orientifolds and
linear/chiral multiplet duality'', Nucl. Phys. B 569 (2000) 362,
hep-th/9910143 ;\\
 C. Srucca and M. Serone, ``Gauge and Gravitational anomalies in
D=4 N=1 orientifolds'', JHEP 9912 (1999) 024
%
%


%``Sigma-model anomalies in compact type IIB orientifolds and
%Fayet-Iliopoulos terms''
%Nucl. Phys. B542 (1999) 112, hep-th/9905098


\bibitem{cim}D. Cremades, L.E. Ibanez, F. Marchesano,
`SUSY Quivers, Intersecting Branes and the Modest Hierarchy Problem',
hep-th/0201205;D. Cremades, L.E. Ibanez, F. Marchesano,'' Intersecting Brane Models of Particle Physics and the Higgs Mechanism'', hep-th/0203160 


\bibitem{senn}A. Sen, JHEP 9808 (1998) 012, 
``Tachyon Condensation on the Brane Antibrane System''
hep-th/9805170; 
``SO(32) Spinors of Type I and Other Solitons on Brane-Antibrane
Pair'', JHEP 809 (1998) 023,
hep-th/9808141.


\bibitem{decay}H. V. Klapdor-Kleingrothaus, A. Dietz, H. L. Harney, I. V.
Krivosheina, ``Evidense for Neutrinoless Double Beta 
decay'', Mod. Phys. Lett. A,
Vol. 16, No 37 (2001)


\bibitem{nsns}
E.~Dudas and J.~Mourad,
``Brane solutions in strings with broken supersymmetry and dilaton
 tadpoles'',
Phys. Lett. B486 (2000) 172, hep-th/0004165; R.~Blumenhagen, A.~Font,
``Dilaton tadpoles, warped geometries and large extra dimensions for
nonsupersymmetric strings'',
Nucl. Phys. B 599 (2001) 241, {\tt hep-th/0011269}.


\bibitem{abla} I. Antoniadis, K. Benakli, A. Laugier, 
``D-brane Models with Non-Linear Supersymmetry'', hep-th/0111209
 

\end{thebibliography}
\end{document}